\documentstyle[psfig]{cupconf}

\ifoldfss
\else
  \ifnfssone
    \newmathalphabet{\mathit}
      \addtoversion{normal}{\mathit}{cmr}{m}{it}
      \addtoversion{bold}{\mathit}{cmr}{bx}{it}
    \newmathalphabet{\mathcal}
      \addtoversion{normal}{\mathcal}{cmsy}{m}{n}
    \else
    \ifnfsstwo
    \fi
  \fi
\fi

\def\hexnumber#1{\ifcase#1 0\or1\or2\or3\or4\or5\or6\or7\or8\or9\or
 A\or B\or C\or D\or E\or F\fi }
\makeatletter
\ifx\CUP@mtlplain@loaded\undefined
\else
\fi
\makeatother
 \makeatletter
 \ifx\CUP@mtlplain@loaded\undefined
   \font\tenbmi=cmmib10 at 10pt
   \font\sevenbmi=cmmib10 at 7pt
   \font\fivebmi=cmmib10 at 5pt

   \newfam\bmifam
   \textfont\bmifam=\tenbmi
   \scriptfont\bmifam=\sevenbmi
   \scriptscriptfont\bmifam=\fivebmi
   
 \fi
 \makeatother
\ifnfsstwo

\fi
\ifnfssone

\fi
\ifoldfss

\fi

\mathchardef\varLambda="0103

\makeatletter
\ifx\CUP@mtlplain@loaded\undefined
\else
\fi
\makeatother
\makeatletter
\ifx\CUP@mtlplain@loaded\undefined
  \font\tenbms=cmbsy10
  \font\sevenbms=cmbsy10 at 7pt
  \font\fivebms=cmbsy10 at 5pt
  \newfam\bmsfam
  \textfont\bmsfam=\tenbms
  \scriptfont\bmsfam=\sevenbms
  \scriptscriptfont\bmsfam=\fivebms

  \edef\bsy@{\hexnumber\bmsfam}
  \mathchardef\bnabla="0\bsy@72
\fi
\makeatother
\def\etal{\mbox{\it et al.}}

\def \apj{\mbox{\it Astrophys. J.}}
\def \apjs{\mbox{\it Astrophys. J. Suppl.}}
\def \mnras{\mbox{\it Mon. Not. R. astr. Soc.}}
\def \aj{\mbox{\em Astron. J.}}
\def \araa{\mbox{\it Ann. Rev. Astron. Astr.}}

\def \nat{\mbox{\it Nature}}
\def \aap{\mbox{\it Astron. Astrophys.}}

\title[Galaxy Formation and Evolution]
{Galaxy Formation and Evolution: Recent Progress}

\author[Richard S. Ellis]{%
R\ls I\ls C\ls H\ls A\ls R\ls D\ns
E\ls L\ls L\ls I\ls S\ns}
  \affiliation{California Institute of Technology MS 105-24,
Pasadena, CA 91125 USA\\[\affilskip]}

\begin{document}
\ifnfssone
\else
  \ifnfsstwo
  \else
    \ifoldfss
      \let\mathcal\cal
      \let\mathrm\rm
      \let\mathsf\sf
    \fi
  \fi
\fi

\maketitle

\begin{abstract}
In this series of lectures\footnote{Lectures given at the XIth Canary
Islands Winter School of Astrophysics ``Galaxies at High Redshift'' in
November 1999 updated to reflect progress in the subject during 2000},
I review recent observational progress in constraining models of galaxy
formation and evolution highlighting the importance advances in
addressing questions of the assembly history and origin of the
Hubble sequence in the context of modern pictures of structure
formation.
\end{abstract}

\firstsection 
\section{Introduction}
\medskip

These are exciting times to be working on any aspect of studies of
galaxies at high redshift whether observational or theoretical.
Most would agree that the current period represents something of a
{\em golden era} in the subject. Figure 1 shows the increasing
extent to which articles concerned with galaxy evolution dominate
the published literature over the past 25 years (gauged
xenophobically I'm afraid by keyword statistics only in two North
American journals).

To try and understand the cause for this prominence in the
subject, the dates associated with the commissioning of some major
observational facilities have been marked. The progress appears to
have been driven largely by new kinds of optical and near-infrared
data: faint counts and searches for primaeval galaxies in the late
1970's and early 1980's (\cite{bap79,tyson79,kron80,koo85}), faint
galaxy redshift surveys made possible by multi-object
spectrographs in the late 1980's and early 1990's
(\cite{lilly95,rse96,cowie96,cohen00}), the launch of Hubble Space
Telescope (HST) and its revelation of resolved galaxy images to
significant redshifts (\cite{griffiths94, glazebrook95,
brinchmann98}), the remarkable Hubble Deep Field image
(\cite{hdf96}) and the plethora of papers that followed
(\cite{hdf98}) and the arrival of the Keck telescopes bringing a
new wave of faint Lyman-break galaxy spectroscopy at unprecedented
redshifts (\cite{steidel96,steidel99})\footnote{A correlation
was also made with three key international conferences
(\cite{yale78,durham88,hdf98}) but I was horrified to see that
these appeared to have had a {\em negative} effect on the
community's output! I assume this arose from a much-needed period
of post-conference reflection!}.

One often hears claims that a subject undergoing spectacular
progress is one that is nearing completion (c.f. \cite{horgan97}).
After all, the rise in Figure 1 clearly cannot continue
indefinitely and fairly soon, it could be argued, we will then
have solved all of the essential problems in the subject. As if
anticipating this, a theoretical colleague gave a recent
colloquium at my institute entitled {\em Galaxy Formation: End of
the Road}\,!

Consider the evidence. Observationally we may soon, via
photometric redshifts, have determined the redshift distribution,
luminosity evolution and spatial clustering of sources to
unprecedented limits. If one accepts photometric redshifts are
reliable, the rate of progress in the traditional pursuit of
$N(m,color, z)$ is limited solely by the field of view of the
telescope and the exposure times adopted. Panchromatic data
matching that obtained with optical and near-infrared telescopes
from SIRTF, FIRST, and ALMA will also enable us unravel the cosmic
star formation history $\rho_{SFR}(z)$ to unprecedented precision
(\cite{madau96,blain99}). It has already been claimed that
the above data, e.g. $N(m,color,z)$ and $\rho_{SFR}(z)$, can be
understood in terms of hierarchical models of structure formation
where galaxies assemble through the cooling of baryonic gas into
merging cold dark matter halos (CDM,
\cite{kauffmann94,baugh98,cole00a}).

The word `concordance' was recently coined astrophysically in an
article reconciling different estimates of the cosmological
parameters (\cite{ostriker96}). Such concordance in our
understanding of galaxy evolution is a natural consequence of
semi-analytical theories whose sole purpose is to explain the `big
picture' as realised with the extant galaxy data. In this series
of lectures I want to show that we have our work cut out for some
considerable time! Exciting progress is definitely being made, but
observers must rise to the challenge of testing the fundamentals
of contemporary theories such as CDM and theorists must get ready
to interpret qualitatively new kinds of data that we can expect in
the next decade.

\bigskip
\centerline{\psfig{file=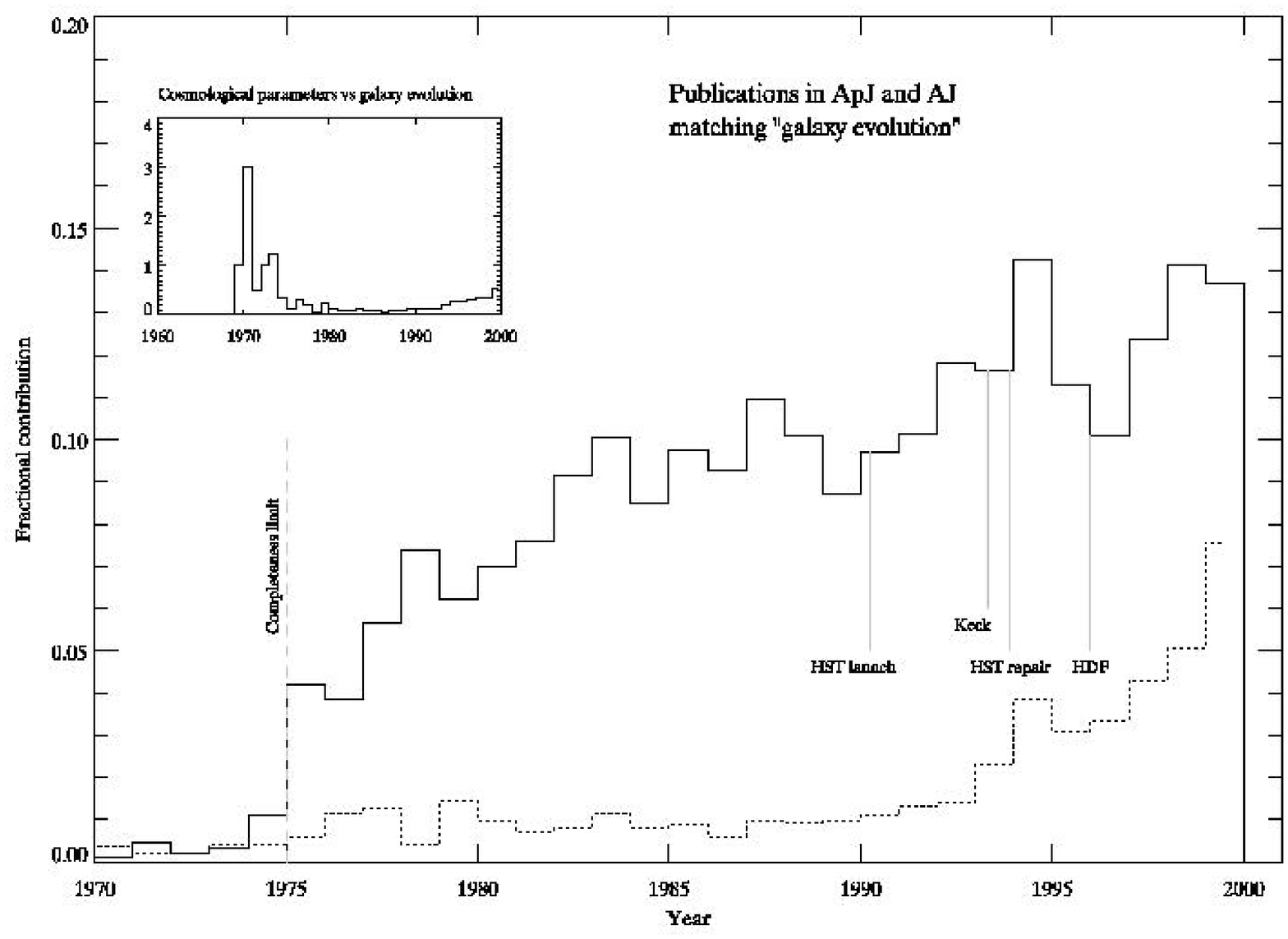,width=9cm,angle=90}}
\noindent{\bf Figure 1:}\quad {\em The remarkably rapid growth in galaxy
evolution studies: the fraction of the ApJ and AJ literature
containing the key word `galaxy evolution' over the past 25 years.
The inset shows the marked decline in the use of galaxies as
probes of the cosmological parameters during 1970-1980 (after
Brinchmann, Ph.D. thesis 1998).}

These lectures are intended for interested graduate students or
postdocs entering the field. There is an obvious observational
flavor although I have tried to keep in perspective an ultimate
goal of comparing results with recent CDM predictions. The bias is
largely to optical and near-infrared applications; there is
insufficient space to do justice to the rapidly-developing
contributions being made at sub-millimetre, radio and X-ray
wavelengths which other contributors at this winter school will
cover in detail.

\bigskip
\section{Galaxy Formation and Cosmology}
\medskip

Traditionally faint galaxies were studied in order to constrain
the cosmological world model (\cite{sandage61}); their evolution
was considered just one more tedious correction (the so-called
{\em evolutionary correction}) in the path to the Holy
Grail of the deceleration parameter $q_o$ ($\equiv \Omega_M$/2 in
$\Lambda$=0 Friedmann models). The most useful galaxies in this
respect were giant ellipticals in rich clusters. Tinsley (1976)
demonstrated how sensitive the derived $q_0$ was to the assumed
main sequence brightening with look-back time in these
populations.

The traditional view for the formation history of an elliptical
followed Eggen, Lynden-Bell \& Sandage (1962). Monolithic collapse and
rapid star formation leads to a subsequent track known as `passive
evolution' (i.e. without further star formation). Tinsley showed that
main sequence brightening in such a stellar population is largely
governed by the rate at which stars evolve off the main sequence, i.e.
the slope $x(\simeq1)$ of the initial mass function at the typical
turnoff mass 0.4--1$M_{\odot}$. Whence:

\begin{equation}
E(z,t) = d\,M_v/d\,lnt \sim 1.3\; - \; 0.3\, x
\end{equation}

and, in terms of its bias on $q_o$:

\begin{equation}
\Delta\,q_o = 1.4 (H_o\,t_o)^{-1}\;d\,M_v/d\,lnt = 1.8\; -\;
0.42\,x
\end{equation}

Tinsley argued that one would have to know the evolutionary
correction to remarkable precision get a secure value of $q_o$. In
fact, noting that the difference in apparent magnitude for a
standard candle at $z$=1 between an empty and Einstein-de Sitter
Universe is only $\simeq$0.5 mag, the relative importance of
cosmology and evolution can be readily gauged.

Despite the above, it is always a mystery to me why several of our
most eminent astronomers (\cite{kristian78,gunn75}) continued to
pursue the Hubble diagram as a cosmological probe using
first-ranked cluster galaxies, in some cases for several years
after the challenge of resolving the evolutionary correction
became known. Tammann (1985) estimated about 400 nights nights of
Palomar 200-inch time was consumed by the two competing groups
whose resulting values of $q_o$ fundamentally disagreed. Recently
Arag\'on-Salamanca (1998) showed, in a elegant summary of the
situation, how the modern $K$-band Hubble diagram is most likely 
complicated further by the fact that first-ranked cluster
galaxies are still assembling their stars over the redshift
interval 0$<z<$1, offsetting the main sequence brightening (Figure
2).

In the late 1970's therefore, the motivation for studying faint
galaxies became one of understanding their history rather than
using them as tracers of the cosmic expansion (see inset panel in
Figure 1). This is not to say that uncertainties in the
cosmological model do not affect the conclusions drawn. The
connection between cosmology and source evolutions remains strong
in three respects:

\medskip
\centerline{\psfig{file=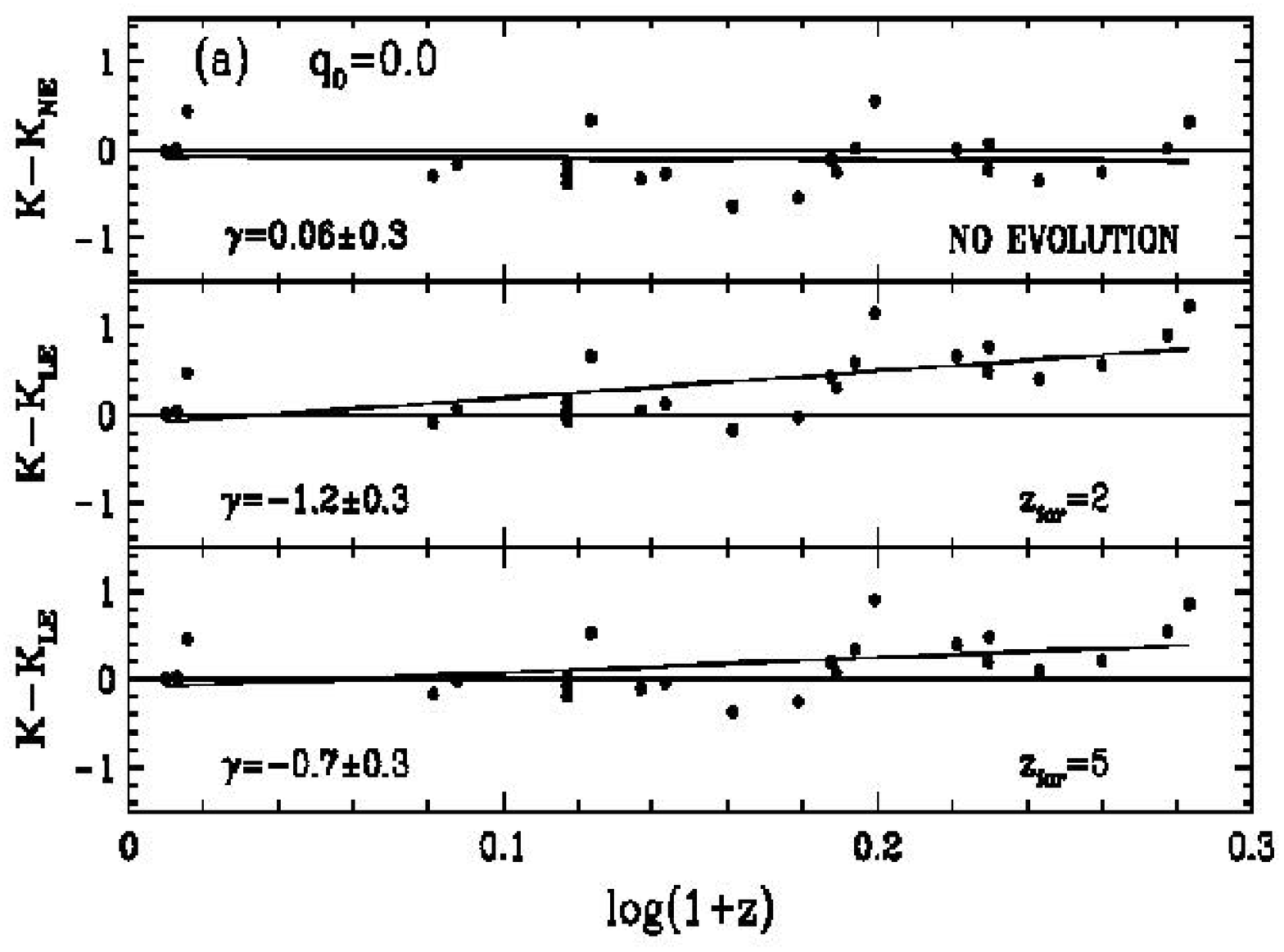,width=11cm,angle=270}}
\noindent{\bf Figure 2:}\quad {\em A recent appraisal of the prospects of
securing cosmological constraints from the Hubble diagram of
brightest cluster galaxies (\cite{aas98}). Luminosity evolution is
parameterised as $L=L(0)\,(1+z)^{\gamma}$. For $q_o$=0, the top
panel shows residuals and best fit trend applying $k$-correction
and luminosity distance effects only; no luminosity evolution is
seen. The middle and bottom panels show the residuals when
evolution is modeled for single burst populations formed at
$z_F$=2 and 5, respectively. High $z$ galaxies are {\em less}
luminous than expected, presumably because they are still
accreting material. Quantitatively, the effect amounts to a factor
of 2-4 less stellar mass depending on the assumed $q_o$ (c.f.
\cite{vandokkum99}).}
\medskip

\medskip

\begin{enumerate}

\item We use our knowledge of stellar evolution to predict the
past appearance of stellar populations in galaxies observed at
high redshift. However, stellar evolution is baselined in {\em
physical time} (the conventional unit is the Gyr: 10$^9$yr),
whereas we observe distant sources in {\em redshift} units. The
mapping of time and redshift depends on the world model. Broadly
speaking there is less time for the necessary changes to occur in
a high $\Omega_M$ universe and consequently evolutionary trends
are much stronger in such models.

\item Many evolutionary tests depend on the {\em numbers} of sources,
the most familiar being the number-magnitude count which is
remarkably sensitive to small changes in source luminosity.
However, the relativistic volume element $dV(z)$ depends
sensitively on curvature being much larger in open and
accelerating Universes than in the Einstein-de Sitter case.

\item Predictions for the mass assembly history of a galaxy in
hierarchical models depend also on the cosmological model in a
fairly complex manner since these models jointly satisfy
constraints concerned with the normalisation of the mass power
spectrum via the present abundance of clusters (e.g.
\cite{baugh98}). Figure 3 illustrates one aspect of this
dependence (\cite{kauffmann98}); structure grows more rapidly in a
dense Universe so the decline with redshift in the abundance of
massive spheroidal galaxies, which are thought in this picture to
forms via mergers of smaller systems, is much more marked in high
density models than in open or accelerating Universes.

\end{enumerate}

\medskip
\centerline{\psfig{file=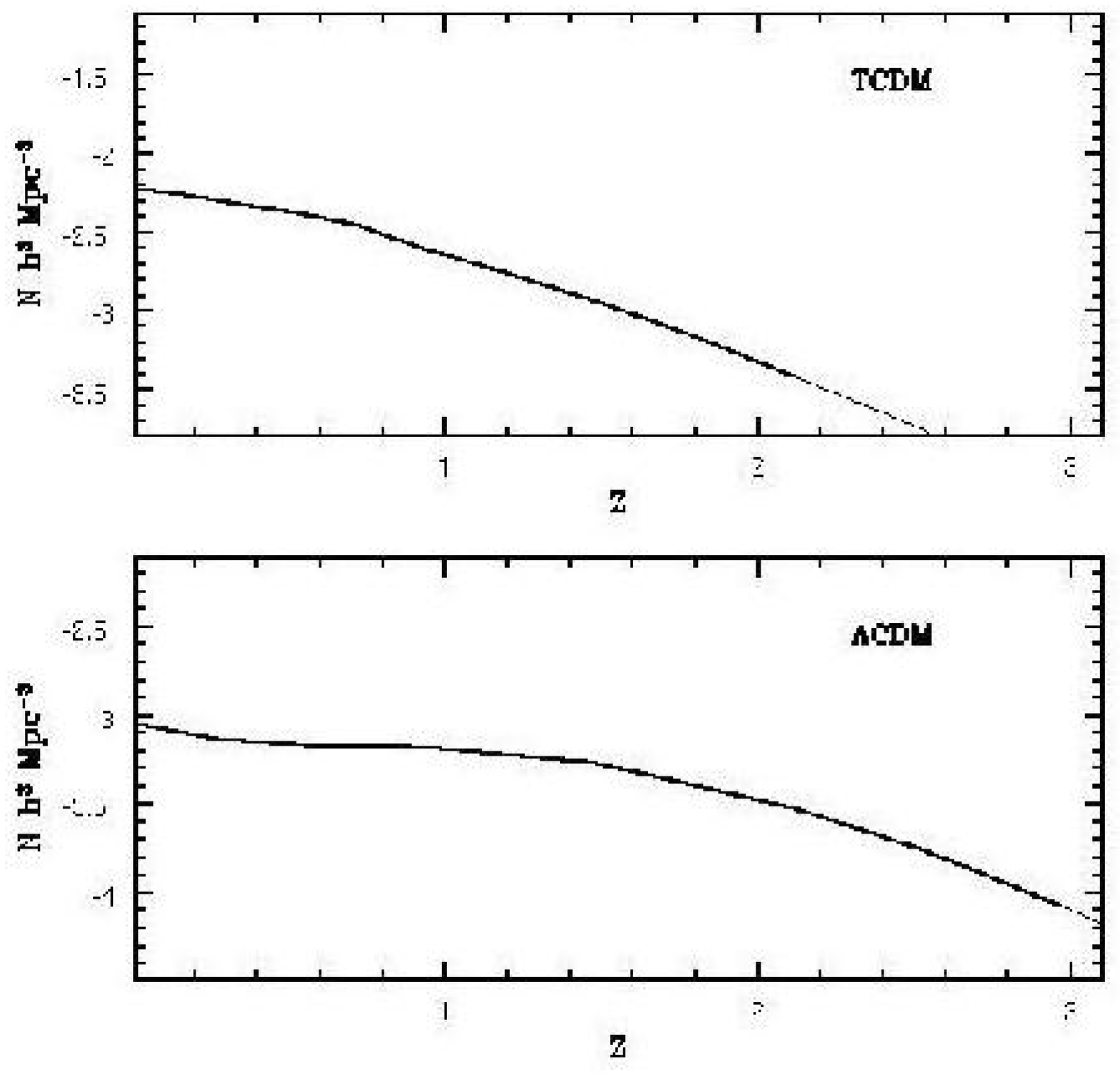,width=8.7cm}}

\noindent{\bf Figure 3:}\quad {\em The abundance of massive
($>$10$^{11}M_{\odot}$) systems as a function of redshift in two
hierarchical models (\cite{kauffmann98}) showing the strong
decline in a high density ($\tau$CDM) model c.f. that in a low
density accelerating model ($\Lambda$CDM)}.
\medskip

Fortunately, we are making excellent progress in constraining the
cosmological parameters from independent methods, the most prominent of
which include the angular fluctuation spectrum in the microwave
background (\cite{de Bernadis00,balbi00}), the Hubble diagram of
distant Type Ia supernovae (\cite{garnavich98,perlmutter99}), the
abundance of rich clusters at various epochs (\cite{bahcall98}) and the
redshift-space distortion in large redshift surveys such as 2dF
(\cite{peacock00}).

Given it matters, how then should we respond to the widely-accepted
{\em concordance} in the determination of $H_o,\Omega_M,\Lambda$ from
various probes (Ostriker \& Steinhardt 1996, Bahcall 1999)? The claimed
convergence on the value of Hubble's constant (\cite{mould00}) is not
so important for the discussion below since most evolutionary tests are
primarily concerned with {\em relative} comparisons at various
look-back times where $H_o$ cancels. The most bewildering aspect of the
concordance picture is the resurrection of a non-zero $\Lambda$, the
evidence for which comes primarily from the Hubble diagram for Type Ia
supernovae.

As a member of the Supernova Cosmology Project (\cite{perlmutter99})
I obviously take the supernova results seriously! However, this
does not prevent me from being surprised as to the implications of
a non-zero $\Lambda$. The most astonishing fact is how readily the
community has apparently accepted the resurrection of $\Lambda$ -
a term for which there is no satisfactory physical explanation
(c.f \cite{wang00}). To one poorly-understand component of the
cosmic energy density (non-baryonic dark matter), we seem to have 
added another (vacuum energy). It seems a remarkable
coincidence that all three significant constituents ($\Omega_B,
\Omega_{DM}, \Omega_{\Lambda}$) are comparable in magnitude to
within a factor of 10, and hardly a step forward that only one 
is physically understood!

The lesson I think we should draw from the {\em cosmic
concordance} is similar to the comment I made in $\S$1 when we
discussed some theorists' triumphant reconciliation of their
theories with faint galaxy data (a point we will debate in detail
in $\S$3). In both cases, the hypothesis certainly reproduces a
wide range of observations but note it takes, as input,
parameters for which there is not yet a clear physical model. One
should not, therefore, regard a concordant picture as anything
other than one of many possible working hypotheses. In the case of
the cosmological models, we need to invest effort into
understanding the physical nature of dark matter and vacuum
energy. In the case of galaxy evolution our goal should be to test
the basic ingredients of hierarchical galaxy formation.

\section{Star Formation Histories}
\medskip

One of the most active areas of relevance to understanding the
rate at which galaxies assemble is concerned with determining the
cosmic star formation history. The idea is simple enough. A
systematic survey is conducted according to some property that is
sensitive to the on-going rate of star formation. The
volume-average luminosity density is converted into its equivalent
star formation rate averaged per unit co-moving volume and the
procedure repeated as a function of redshift to give the cosmic
star formation history $\rho_{\ast}(z)$. In this section we will
explore the uncertainties and also the significance of this
considerable area of current activity in terms of the constraints
they provide on theories of galaxy formation.

The joint distribution of luminosity $L$ and redshift $z$,
$N(L,z)$, for a flux-limited sample permits the construction of
the luminosity function $\Phi(L)$ according to procedures which
are reviewed by Efstathiou, Ellis \& Peterson (1988) and compared
by Ellis (1997) . The luminosity function is often characterised
according to the form defined by Schechter (1976), viz:

\begin{equation}
\Phi(L)\,dL/L^{\ast} =
\Phi^{\ast}\,(L/L^{\ast})^{-\alpha}\,exp(-L/L^{\ast})\,dL/L^{\ast}
\end{equation}
in which case the integrated number of galaxies per unit volume
$N$ and the luminosity density $\rho_L$ then becomes:
\begin{equation}
N= \int{\Phi(L) \, dL} = \Phi^{\ast}\,\Gamma(\alpha+1)
\end{equation}
and
\begin{equation}
\rho_L= \int{\Phi(L)\, L\, dL} =
\Phi^{\ast}\,L^{\ast}\,\Gamma(\alpha+2)
\end{equation}
and the source counts in the non-relativistic case, applicable to
local catalogs, is:
\begin{equation}
N(<m)\; \propto \; d^{\ast\,3}(m) \int
{dL\,\Phi(L)\,(L/L^{\ast})^{{\frac{3}{2}}}} \;\propto \;
\Phi^{\ast}\,L^{\ast\,\frac{3}{2}}\,\Gamma(\alpha\,+\,\frac{5}{2})
\end{equation}

Frequently-used measures of star formation in galaxies over a range of
redshift include rest-frame ultraviolet and blue broad-band
luminosities (\cite{lilly95,steidel96,sullivan00}), nebular emission
lines such as H$\alpha$ (Gallego et al 1995, Tresse \& Maddox 1998,
Glazebrook et al 1999), thermal far-infrared emission from dust clouds
(\cite{mrr97,blain99}) and, most recently, radio continuum emission
(\cite{mobasher99}).

Since only a limited range of the luminosity function centered on
$L^{\ast}$ is reliably probed in flux-limited samples, a key issue
is how well the integrated luminosity density $\rho_L$ can be
determined from such surveys. In the Schechter formalism,
equations [3.4] and [3.5] show that whilst $N$ would diverge for
$\alpha<$--1, the luminosity density is convergent unless
$\alpha<$--2.

\medskip
\centerline{\psfig{file=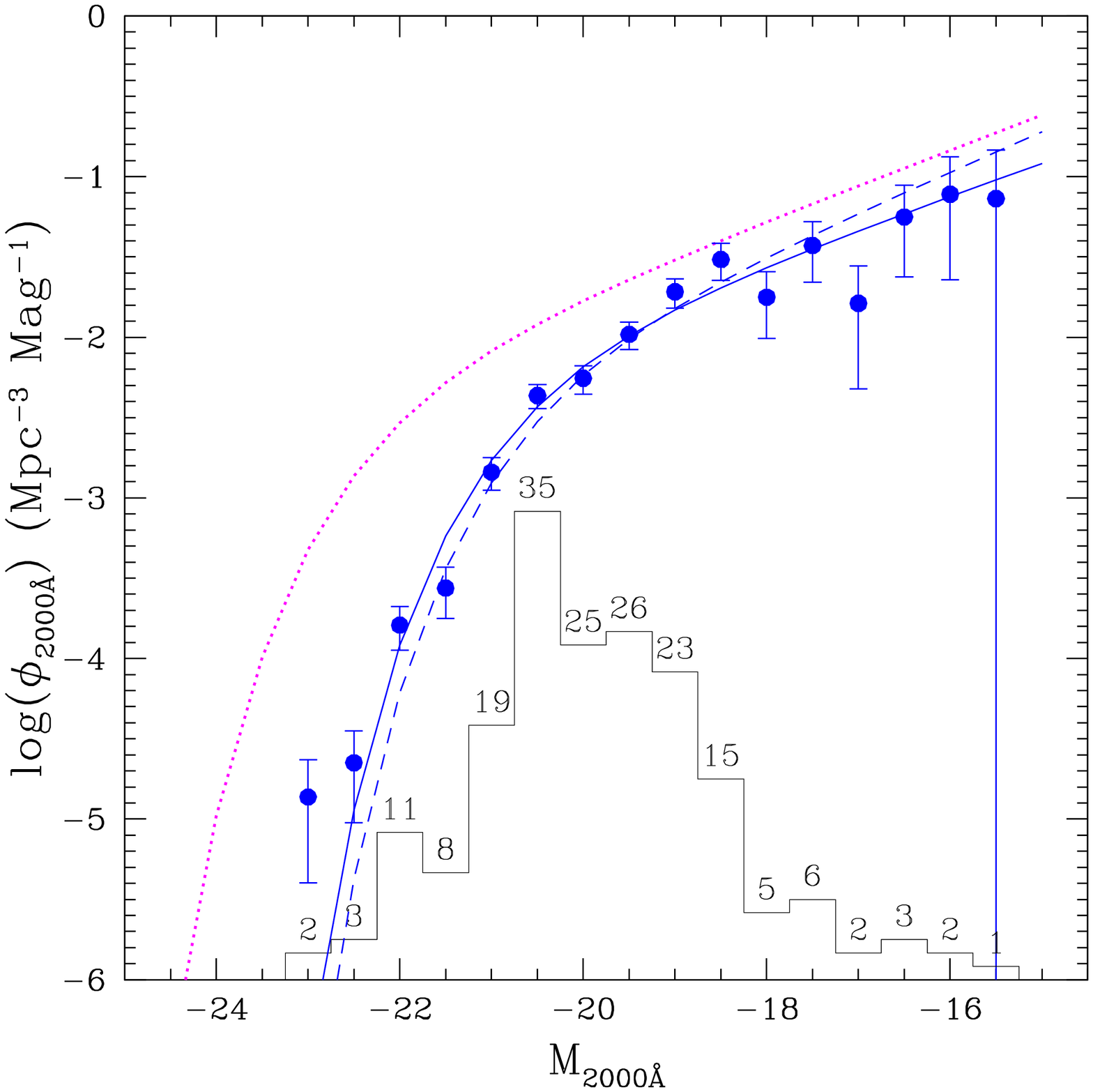,width=9cm}}
\noindent{\bf Figure 4:}\quad {\em The luminosity function for galaxies
selected at 2000 \AA\ from the recent survey of Sullivan et al
(2000). The histogram and associated numbers indicate the absolute
magnitude distribution observed which is corrected by volume and
$k$-correction effects to give the data points. The dotted curve
illustrates the considerable effect of extinction as gauged by
Balmer decrements determined individually for those galaxies with
emission lines. Such uncertainties translate in factors of two
uncertainty in the local UV luminosity density.}
\medskip

Figure 4 shows the local rest-frame ultraviolet (2000 \AA\ )
luminosity function from Sullivan et al (2000) whose faint end
slope $\alpha$=--1.6 is markedly steeper than that found for
samples selected in the near-infrared
(\cite{mobasher93,gardner97,cole00b}) (where $\alpha\simeq$-1).
This contrast in the luminosity distribution of young and old
stellar populations is an important result which emphasizes the
relatively weak connection between stellar mass and light and
implies there may be significant uncertainties in the estimation
of integrated luminosity densities for star-forming populations.

Kennicutt (1998) carefully reviewed the relationships between the
various observational diagnostics listed above and the star
formation rate. Clearly a major uncertainty in any transformation
based on the ultraviolet/optical continuum or nebular emission
line measures is the likely presence of absorbing dust (Figure 4).
Other uncertainties include the form of the initial stellar mass
function and the nature of the star formation history itself.

\bigskip

\centerline{\psfig{file=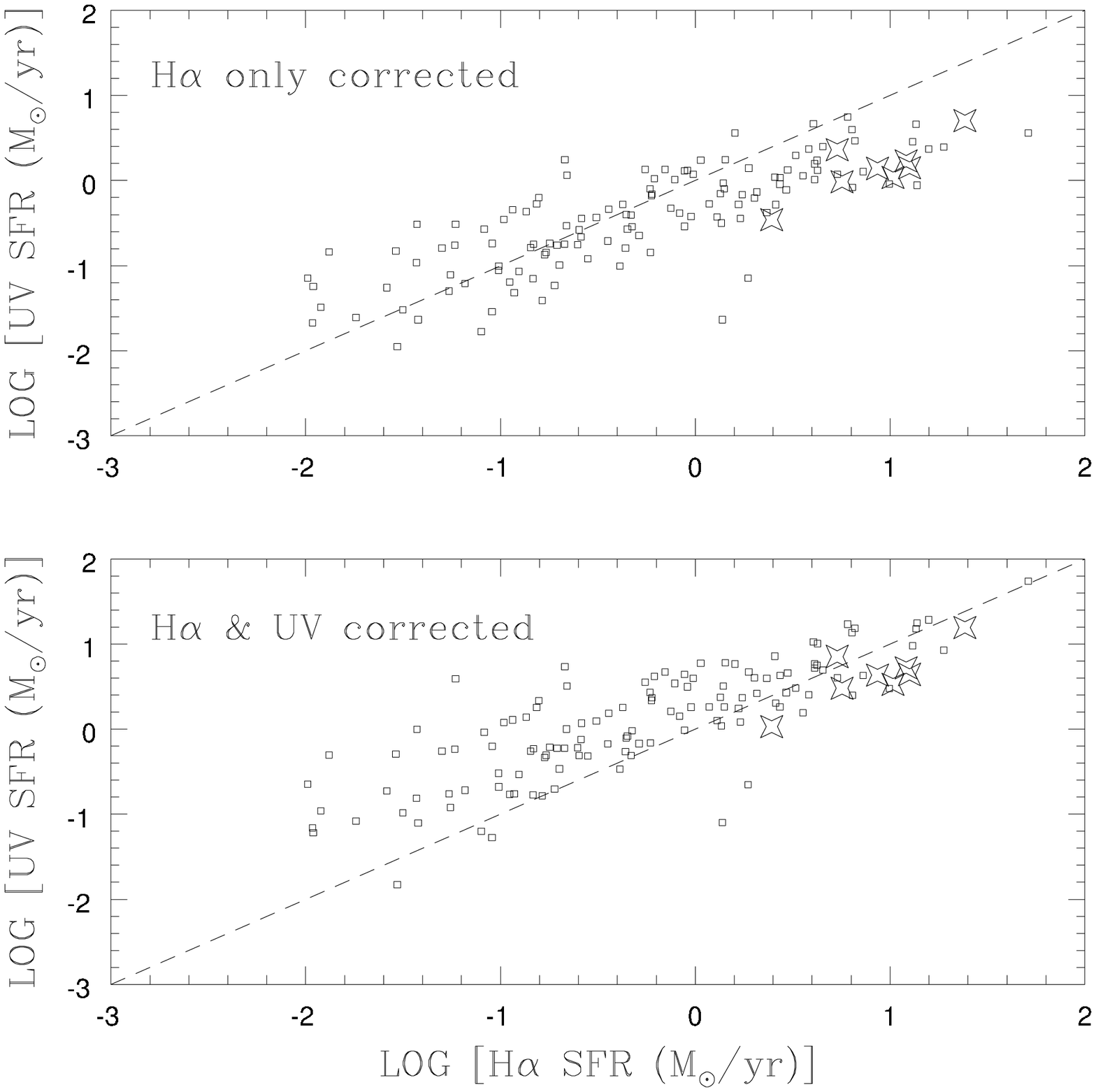,width=9cm}}
\noindent{\bf Figure 5:}\quad {\em Star formation rates derived from UV
(2000 \AA\ ) continua versus those derived from H$\alpha$ fluxes
from the local survey of Sullivan et al (2000, open squares) and
the $z\simeq$1 samples of Glazebrook et al (1999, large stars).
For the Sullivan et al sample, extinction corrections were derived
from individual Balmer decrements assuming Case B recombination
and applied to the $H\alpha$ fluxes in the upper panel and both
estimates in the lower panel.}
\medskip

Some of these uncertainties are quite imponderable and the only
way to estimate their effect in typical populations is to
undertake a comparison of the various diagnostics for the same
sample. Sullivan et al (2000) compared UV and H$\alpha$-based
estimators for their local balloon-based UV-selected sample and
Glazebrook et al (1999) undertook a similar comparison for a
restricted incomplete sample of high redshift galaxies (drawn from
a $I$-selected sample). Bell \& Kennicutt (2000) independently
examined some of Sullivan et al's conclusions based on a smaller
local sample with satellite UV fluxes. The comparison analysed by
Sullivan et al is shown in Figure 5. Although an overall linear
relation is observed the scatter is quite considerable, greater
than accountable from observational errors. The uncertainties
would appear to be alarming in view of the fairly modest trends
claimed in $\rho_{SFR}(z)$ (see below).

In addition to the scatter arising from extinction (accounted for
via individual Balmer emission line decrements), Sullivan et al
suggest that some fraction of their UV-selected population must be
suffering star formation which is erratic in its time history. In
such a situation, different diagnostics will be sensitive to
bursts of activity for different periods, corresponding to the
time over which the contributing stars remain on the main
sequence. $H\alpha$ flux arises from recombination photons linked
to those emitted below the Lyman limit from main sequence stars
with lifetimes $\simeq$10$^6$ years. The UV and blue continua
persist for much longer periods ($\simeq 10^8 - 10^9$ years).

Depending upon how widespread star formation histories of this
kind may be, two forms of error may arise in estimating cosmic
star formation histories. Firstly, the star formation rate derived
for an individual galaxy will be a past time average, smoothing
over any erratic behavior, rather than a true instantaneous value.
More importantly however, particularly at high redshift, galaxies
may be preferentially selected only if their star formation
history is erratic, for example in $H\alpha$ surveys where some
threshold of detectability may seriously restrict the samples.

Figure 6 shows a recent estimate of the cosmic star formation
history drawn from various surveys (\cite{blain00}). There appears
to be a marked increase in activity over 0$<z<$1 with a possible
decline beyond $z>$2. Although, inevitably perhaps, attention has
focused on the case for the high redshift decline, even the strong
rise to $z\simeq$1 remains controversial. Originally proposed
independently by Lilly et al (1995) and Fall et al (1996), revised
estimates for the local luminosity density (\cite{sullivan00}) and
independent surveys (\cite{cowie99}) have challenged the rapidity
of this rise. Part of the problem is that no single survey permits
a self-consistent measurement of $\rho_{SFR}$ over more than a
very limited range in $z$. Most likely, therefore, much of the
scatter in Figure 6 is simply a manefestation of the kinds of
uncertainties discussed above in the context of Sullivan et al's
survey.

Beyond z$\simeq$2, the available star formation rates have been
derived almost exclusively from UV continua in Lyman break
galaxies selected by their `dropout' signatures in various
photometric bands (\cite{madau96,steidel96,steidel99}) and from
currently scant datasets of sub-mm sources interpreted assuming
thermal emission from dust heated by young stars
(\cite{blain99,barger99b}). There has been much discussion on the
possible disparity between the estimates derived from these two
diagnostics (which other lecturers will address). Two points can
be made: firstly, the measured UV luminosity densities will
clearly underestimate the true values given likely extinctions.
Secondly, the sample of sub-mm sources with reliable redshifts
remains quite inadequate for luminosity density estimates in the
sense described above. Most of the constraints arise from
modelling their likely properties in a manner consistent with
their source counts and the integrated far-infrared background.

Have we become over-obsessed with determining the cosmic star
formation history? Observers are eager to place their survey
points alongside others on the overall curve and different groups
defend their methods against those whose data points disagree. We
should consider carefully what role this cosmic star formation
history plays in understanding how galaxies form?

\bigskip
\centerline{\psfig{file=iac_fig6.ps,width=10cm,angle=270}}

\noindent{\bf Figure 6:}\quad {\em The history of recent star formation
from the recent compilation of Blain (2000). Data points are taken
from a variety of sources referenced in that article. Thick solid
and dashed lines represent trends expected from simple luminosity
evolution and hierarchical models, respectively. It is clear there
is considerable observational scatter at all redshifts, not just
beyond $z\simeq$1 as often assumed.}
\medskip

Clearly, the prime conclusion we can draw from Figure 6 is that
the stars which make the galaxies we see today formed continuously
over a very wide redshift range. This may seem such an obvious
deduction that it hardly merits stating but it is important to
stress the absence of any obvious detectable `epoch of star
formation' as was once imagined (\cite{els62,durham88}).
Hierarchical modelers were quick to point out (e.g.
\cite{baugh98}) that they predicted extended star formation
histories as early as 1990 (\cite{white91}). It is certainly true
that a continuous assembly of galaxies is a major feature of these
models and thus one supported by the data.

However, what about the {\em quantitative} form of Figure 6 which
remains so difficult to pin down: does the shape of the curve
really matter? Firstly, we should recognise that the luminosity
density integrates over much detailed astrophysics that may be
important. A particular $\rho_{SFR}$ at a given redshift could be
consistent {\em either} with a population of established massive
sources undergoing modest continous star formation {\em or} a
steep luminosity function where most of the activity is in
newly-formed dwarf galaxies. In terms of structure formation
theories, these are very different physical situations yet that
distinction is lost in Figure 6.

Secondly, theoretically, the cosmic star formation history is not
particularly closely related to how galaxies assemble. It is more
sensitive to the rate at which gas cools into the assembling dark
matter halos, a process of considerable interest but which
involves a myriad of uncertain astrophysical processes (Figure 7)
which are fairly detached from the underlying physical basis of
say the hierarchical picture. In support of this, we should note
that Baugh et al (1999) were able, within the same
$\Lambda$-dominated CDM model, to `refine' their earlier prediction
to match new high redshift datapoints revealing a much less marked
decline beyond z$\simeq$2.

\bigskip

\centerline{\psfig{file=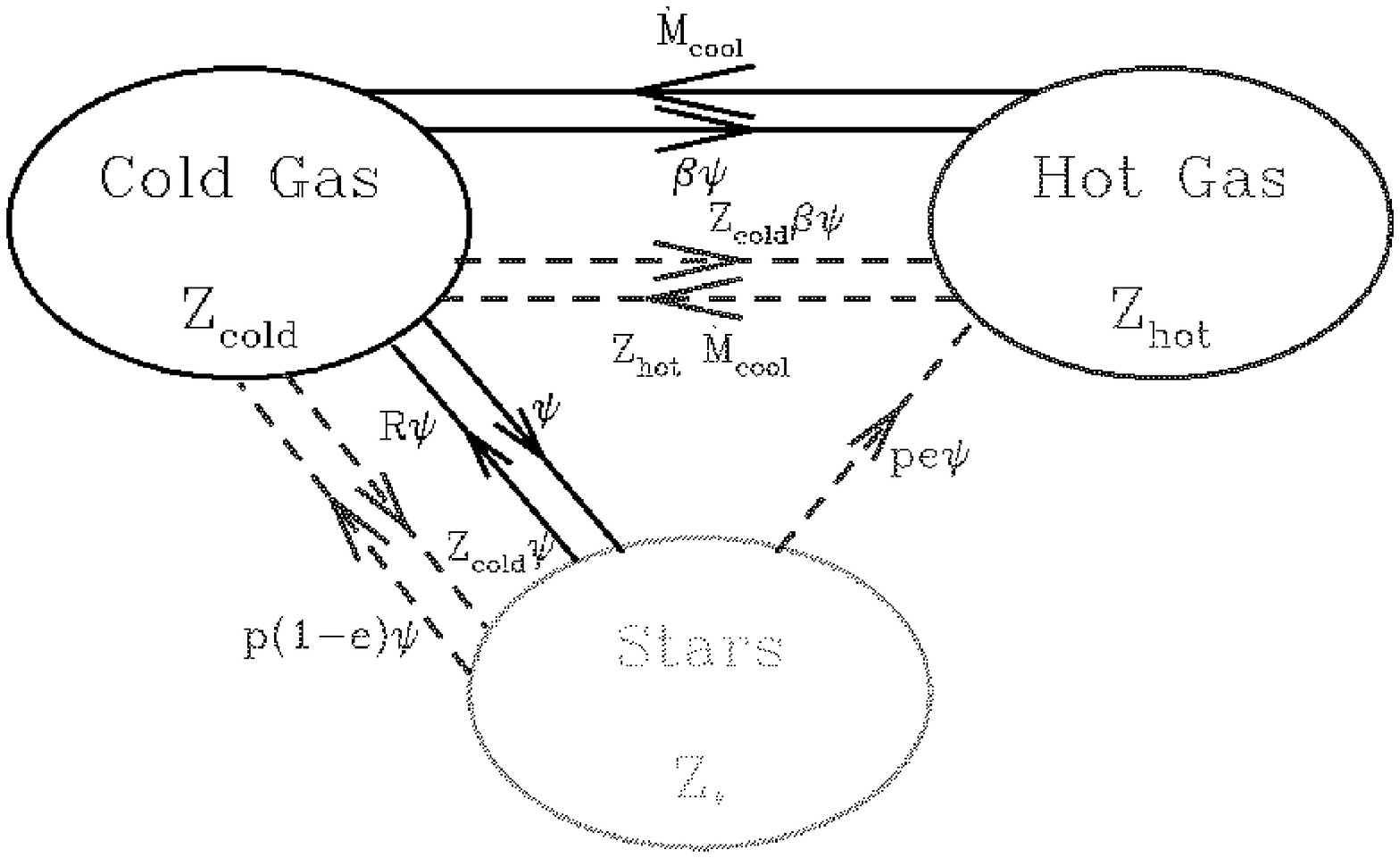,width=10cm}}

\noindent{\bf Figure 7:}\quad {\em An illustration of the complex physical
processes governing the star formation rate of a young galaxy
(courtesy of Carlos Frenk). Star formation is governed by the rate
at which baryonic gas cools and falls into dark matter halos and
this is inhibited by heating, e.g. from supernovae. The precise
form of the cosmic star formation history gives us more insight
into the interplay between these processes, integrated over all
star-forming galaxies, than in distinguishing between various
forms of structure formation (e.g. hierarchical vs. monolithic).}

\bigskip

\section{Morphological Data from HST}
\medskip

As we discussed in $\S$1, one of the most exciting new datasets
that arrived in the mid-1990's was the first set of resolved
images of galaxies at significant look-back times from HST. Much
of the early work was conducted in rich clusters
(\cite{couch94,dressler94,couch98,dressler98}) where the
well-known `Butcher-Oemler' effect (\cite{bo78}) - a surprisingly
recent increase in the fraction of blue cluster members - was
found to be due to a dramatic shift in the morphology-density
relation (Figure 8). As recently as $z\simeq$0.3-0.4 (3-4 Gyr
ago), cluster S0s were noticeably fewer in proportion, their place
apparently taken by spirals, many of which showed signs of recent
disturbances, such a distorted arms and tidal tails.

\medskip
\begin{figure}
\centerline{\hbox{
\psfig{file=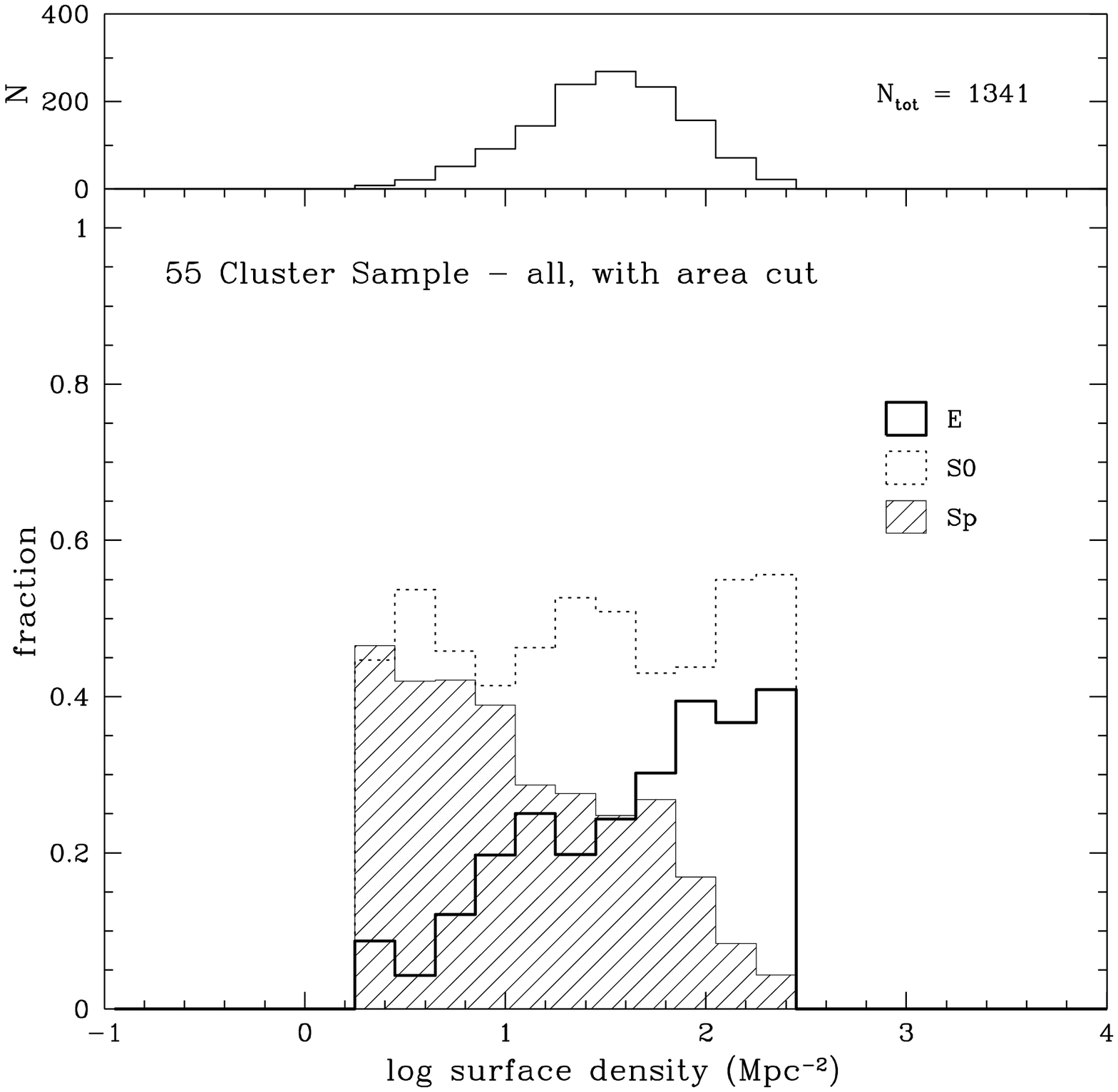,width=6cm}    
\psfig{file=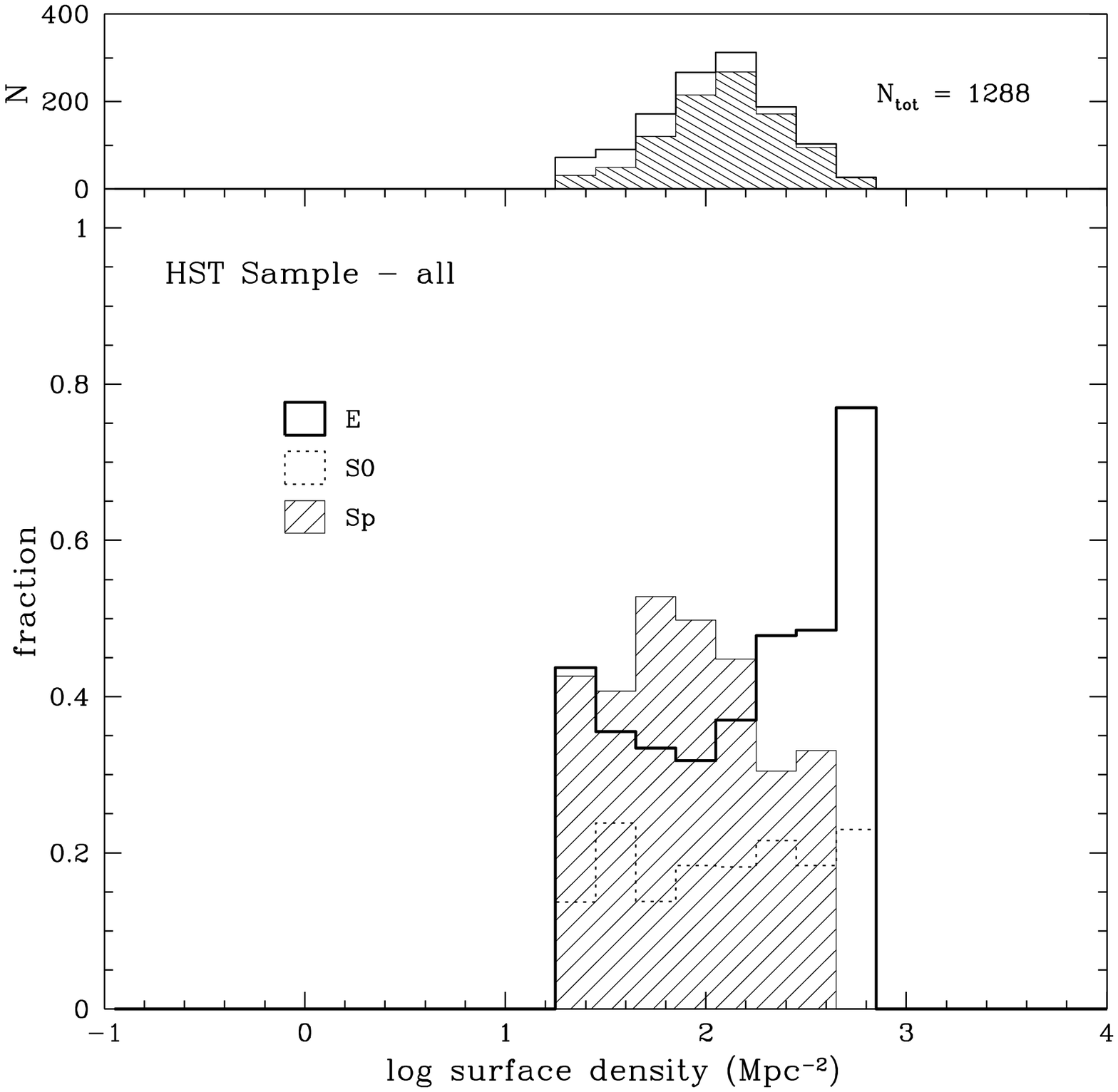,width=6cm}
}}
\noindent{\bf Figure 8:}\quad {\em Evolution in the morphology density
relation from the `Morphs' survey of Dressler et al (1998). (Left)
The fraction of E/S0/Spirals as a function of projected galaxian
surface density for Dressler's 55 local cluster sample. (Right) As
left, for all distant HST clusters with 0.3$<z<$0.55. The
comparison refers to the same cluster core radius
($<$0.6Mpc,$h$=0.5) and includes galaxies to the same rest-frame
$V$ luminosity ($M_V$=-20.0). Note the dramatic decline in the S0
population and the marked increase in the spiral fraction for
environments with high projected density.}
\label{}
\end{figure}
\medskip

The physical origin of this transformation from spirals to S0s
remains unclear and is currently being explored by detailed
spectroscopy of representative cluster members
(\cite{barger96,abraham96b,poggianti99}). A key diagnostic here is
the interplay between the changing morphologies, the presence of
nebular emission lines (such as [O II] 3727 \AA\ -- H$\alpha$ is
generally redshifted out of the accessible range) and Balmer
absorption lines (such as H$\delta$ 4101 \AA\ ). The latter lines
are prominent in main sequence A stars which linger for $\simeq$1
Gyr after any enhanced starburst activity.

Barger et al (1996) proposed a simple cycle where an unsuspecting
galaxy undergoes some perturbation, perhaps due to a merger or its
first encounter with the intracluster gas, subsequently becomes
morphologically-distorted and spectrally-active before subsiding
to a regular spheroidal with a decaying Balmer absorption line.
Whereas such a cycle can explain the {\em proportion} of unusual
objects, it has difficulty matching their {\em luminosities}. A
galaxy should be rendered more luminous during a burst and thus
blue examples cannot easily be the precursors of the
equally-luminous red {\em post-burst} cases. A controversy has
since arisen over the fractions of objects seen in the various
spectrally-active classes (\cite{balogh99}) suggesting much work
is needed in this area, both in quantifying cluster-cluster
variations and also radial variations in the responsible
processes.

Although the cluster work discussed above represents something of
a digression in our overall theme, the realisation that galaxies
can so easily be transformed morphologically has profound
implications for our understanding of galaxy formation. Much 
of the early work explaining the Hubble sequence
(\cite{tinsley77}) assumed galaxies evolve as isolated systems,
however the abundance of morphologically-peculiar and interacting galaxies
in early HST images (\cite{griffiths94}) has been used to
emphasize the important role that galaxy mergers must play in
shaping the present Hubble sequence (\cite{toomre72,barnes92}).
Merger-induced transformations of this kind are a natural
consequence of hierarchical models (\cite{baugh96}). Early disk
systems are prone to merge during epochs when the cosmic density
is high and the peculiar velocity field is cold, forming
bulge-dominated and spheroidal systems which may then later
accrete disks.

The possibility that galaxies transform from one class to another is a
hard hypothesis to verify observationally since, as we have seen,
traditionally observers have searched for redshift-dependent trends
with subsets of the population chosen via an observed property (color,
morphology, spectral characteristics) which could be transient.
Moreover, experience ought to teach us that the outcome of tests of
galaxy formation rarely come down simply to either Theory A {\em or}
Theory B; usually it is some complicated mixture or the question was
naive in the first place! Fortunately, the late formation of massive
regular galaxies in the hierarchical picture (Figure 9) seems a
particularly robust prediction and one in stark contrast to the
classical `monolithic collapse' picture (\cite{tinsley77,sandage83}).
The distinction is greatest for ellipticals presumed to form at high
redshift with minimum dissipation (their central density reflecting
that of the epoch of formation). Studying the evolutionary history of
massive ellipticals is thus an obvious place to start.

\medskip
\centerline{\psfig{file=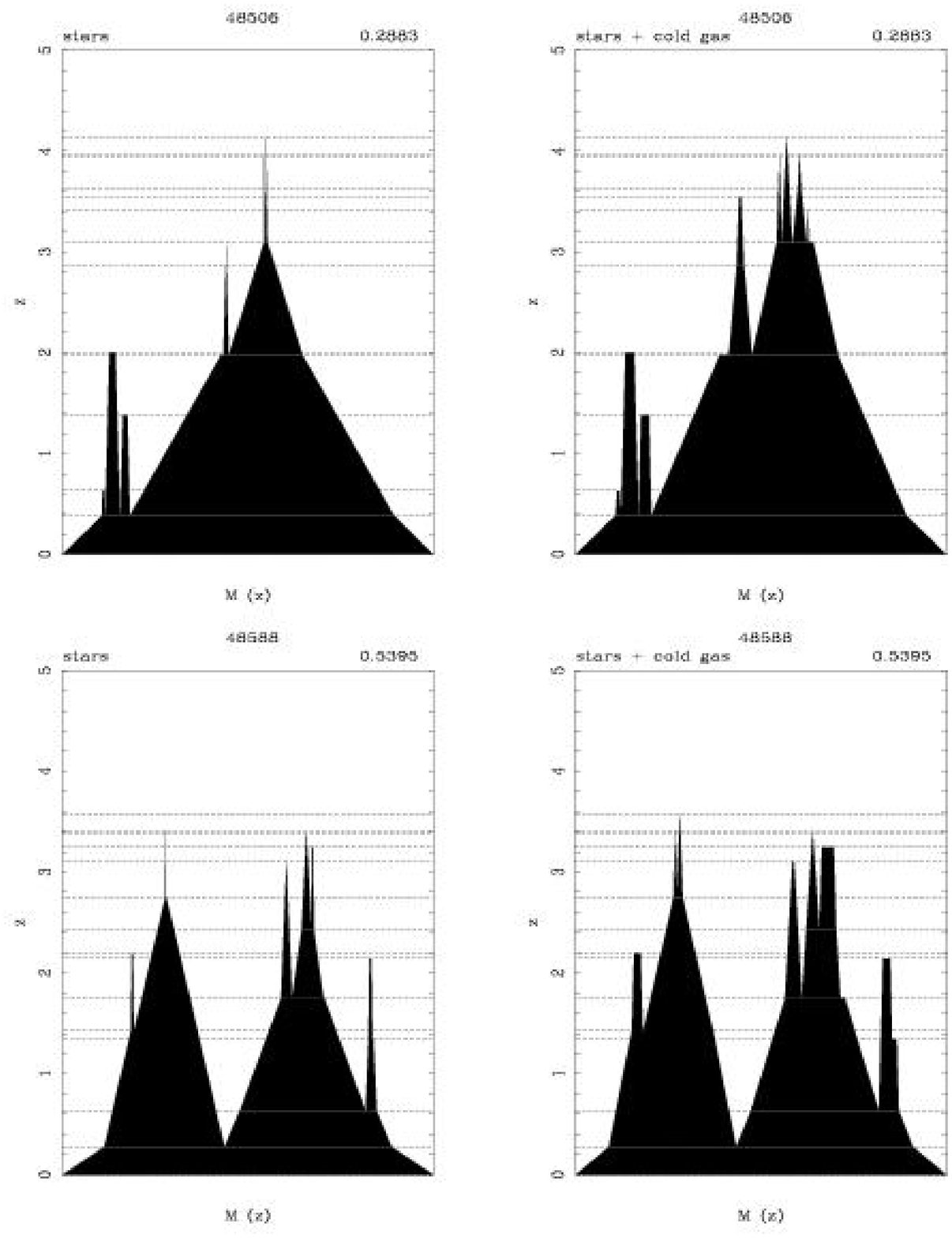,width=8cm}}

\noindent{\bf Figure 9:}\quad {\em The important role of late merging in a
typical CDM semi-analytical model (Baugh et al 1996). The panels
show the redshift-dependent growth for the stellar mass (left) and
that of all baryonic material (right), as indicated by the
thickness of the black area at a given epoch, for two present-day
massive galaxies. The top system grows gradually and is thought to
represent a present-day spiral. The bottom system suffers a late
equal-mass merger thought to produce a present-day elliptical.
Note the remarkably late assembly; most of stellar mass in both
cases assembles in the interval 0$<z<$1. }
\medskip

An oft-quoted result in support of old ellipticals is the
remarkable homogeneity of their optical colors (\cite{sv78,
bower92}). The idea is simple: the intrinsic population scatter in
a color sensitive to recent star formation, such as $U-B$, places
a constraint either on how synchronous the previous star formation
history must have been across the population or, if galaxies form
independently, the mean age of their stellar populations. By
combining cluster data at low redshift (\cite{bower92}) with
HST-selected samples at intermediate redshift (\cite{rse98}), the
bulk of the cluster elliptical population was deduced to have
formed its stars before z$\simeq$2, in apparent conflict with
hierarchical models. Similar conclusions have been drawn from
evolution of the mass/light ratio deduced from the fundamental
plane (\cite{ziegler97,vandokkum98}).

Unfortunately, one cannot generalize from the results found in
distant clusters. In hierarchical models, clusters represent early
peaks in the density fluctuations and thus evolution is likely
accelerated in these environments (\cite{kauffmann95}) plus, of
course, there may be processes peculiar to these environments
involving the intracluster gas. It is also important to
distinguish between the history of {\em mass} assembly and that of
the {\em stars}. Recent evidence for widespread merging of
ellipticals in clusters (\cite{vandokkum99}) lends support to the
idea that the stars in dense regions were formed at high redshift,
in lower mass systems which later merged.

For these reasons, attention has recently switched to tracking the
evolution of {\em field} ellipticals. The term {\em field
elliptical} is something of a misnomer here since a high fraction
of ellipticals actually reside in clusters. What is really meant
in this case is that we prefer to select ellipticals
systematically in flux-limited samples rather than concentrate on
those found in the cores of dense clusters\footnote{A major
concern in all the work relating to the evolution of galaxies in
clusters is precisely how the clusters were located.}.

The study of evolution in field ellipticals is currently very
active and I cannot possibly do justice, in the space available,
to the many complex issues being discussed. Instead let me
summarise what I think are the most interesting results.

\medskip
\begin{itemize}

\item Searches for a population of faint intrinsically red objects,
representing the expected $z>$1 precursors of passively-evolving
ellipticals which formed their stars at high redshift have been
conducted both with and without HST morphological data
(\cite{zepf97,barger99a,menanteau99,daddi00,mccarthy00}). However,
only recently have substantial areas of sky been mapped. This is
because such searches are most sensitive to high $z$ sources when
conducted using optical-near infrared colors and access to large
infrared arrays is a recent technical development. Both Daddi et
al (2000) and McCarthy et al (2000) (Figure 10) claim strong
angular clustering in their faint red populations and there is
limited evidence that the abundance is consistent with a constant
comoving number density, in contrast to the hierarchical
predictions. However, without confirmatory spectroscopy neither
the redshift range nor the nature of these red sources is yet
clear. Even if it later emerges, as was claimed on far less
convincing data (\cite{zepf97,barger99a,menanteau99}), that there
is a shortage of intrinsically red objects beyond $z\simeq$1, only
a modest amount of residual star formation is needed to
substantially bluen a well-established old galaxy
(\cite{jimenez99}). Even when redshift data is secured, color alone
may be an unreliable way to track a specific population.

\medskip
\centerline{\psfig{file=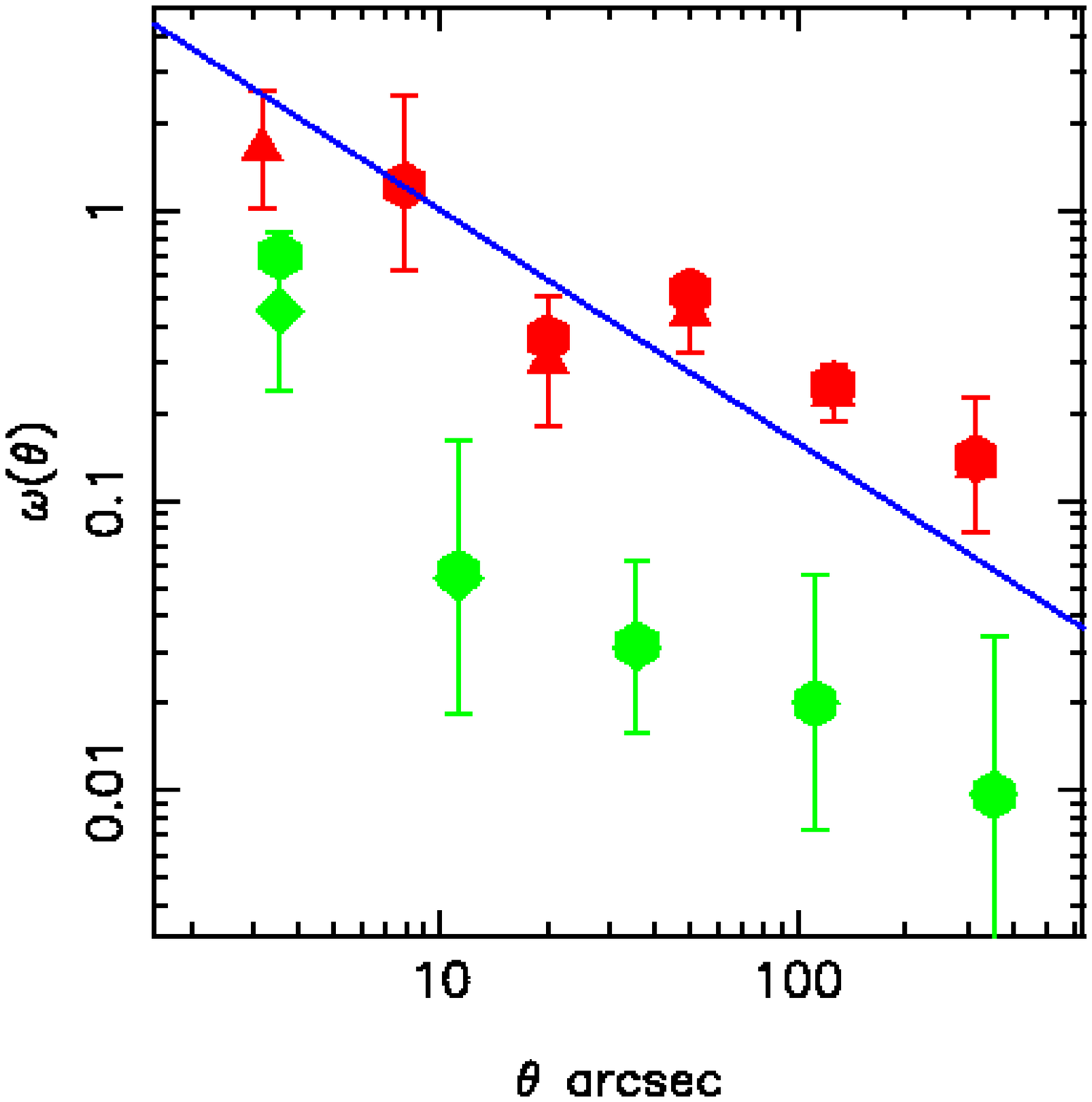,width=7cm}}

\noindent{\bf Figure 10:}\quad {\em Evidence for a clustered population of
red objects in a 1000 arcmin$^2$ area of the ongoing Las Campanas
Infrared Survey (\cite{mccarthy00}). The angular correlation
function for objects with $I-H>$3.5 (top set of data points) is
substantially above that for all $H$-selected galaxies (bottom
set) and consistent with a high fraction of the red objects being
clustered ellipticals with $\overline{z}\simeq$1.0-1.5.}
\medskip

\item At brighter magnitudes, systematic redshift surveys
with associated HST data give constraints on the luminosity
function and colors of morphologically-selected ellipticals
(\cite{brinchmann98,schade99,menanteau99,im00}). Unfortunately,
because of the disparity in field of view betweeen WFPC-2 and the
ground-based multislit spectrographs, the samples remain small and
hence the conclusions are subject to significant field-to-field
clustering uncertainties. However, no substantial decline in the
volume density of ellipticals is yet observed to $z\simeq$1,
although there is some dispute as to the fraction which may
deviate in color from the passive track (\cite{im00} c.f.
\cite{menanteau99}). Current spectroscopic surveys may not be
quite deep enough to critically test the expected evolution in the
hierarchical models, particularly if $\Lambda\neq$0.

\item A completely independent method of determining whether field
ellipticals form continously as expected in hierarchical models is
possible in the Hubble Deep Fields (Menanteau et al 2000). Here, the
imaging signal/noise is sufficient to permit an examination of the
{\em internal} colors of ellipticals with $I<$24, a subset of
which have redshifts. Menanteau et al (2000) find about 25\% of
the HDF ellipticals show blue cores and other color
inhomogeneities suggestive of recent star formation, perhaps as a
result of the merger with a gas-rich low mass galaxy. Keck
spectroscopy (\cite{ellis01}) supports this suggestion: galaxies
with blue cores generally show emission and absorption line
features indicative of star formation (Figure 11). The amount of
blue light seen in the affected ellipticals can be used to
quantify the {\em amount} of recent star formation and the
associated spectrum can be used to estimate the {\em timescale} of
activity through diagnostic features of main sequence stars. Only
modest accretion rates of $\simeq$10\% by mass over $\simeq$1 Gyr
are implied, albeit for a significant fraction of the population.
This continued growth, whilst modest c.f. expectations of
hierarchical models, is noticeably {\em less} prominent in rich
clusters (\cite{menanteau00}).

\medskip
\centerline{\psfig{file=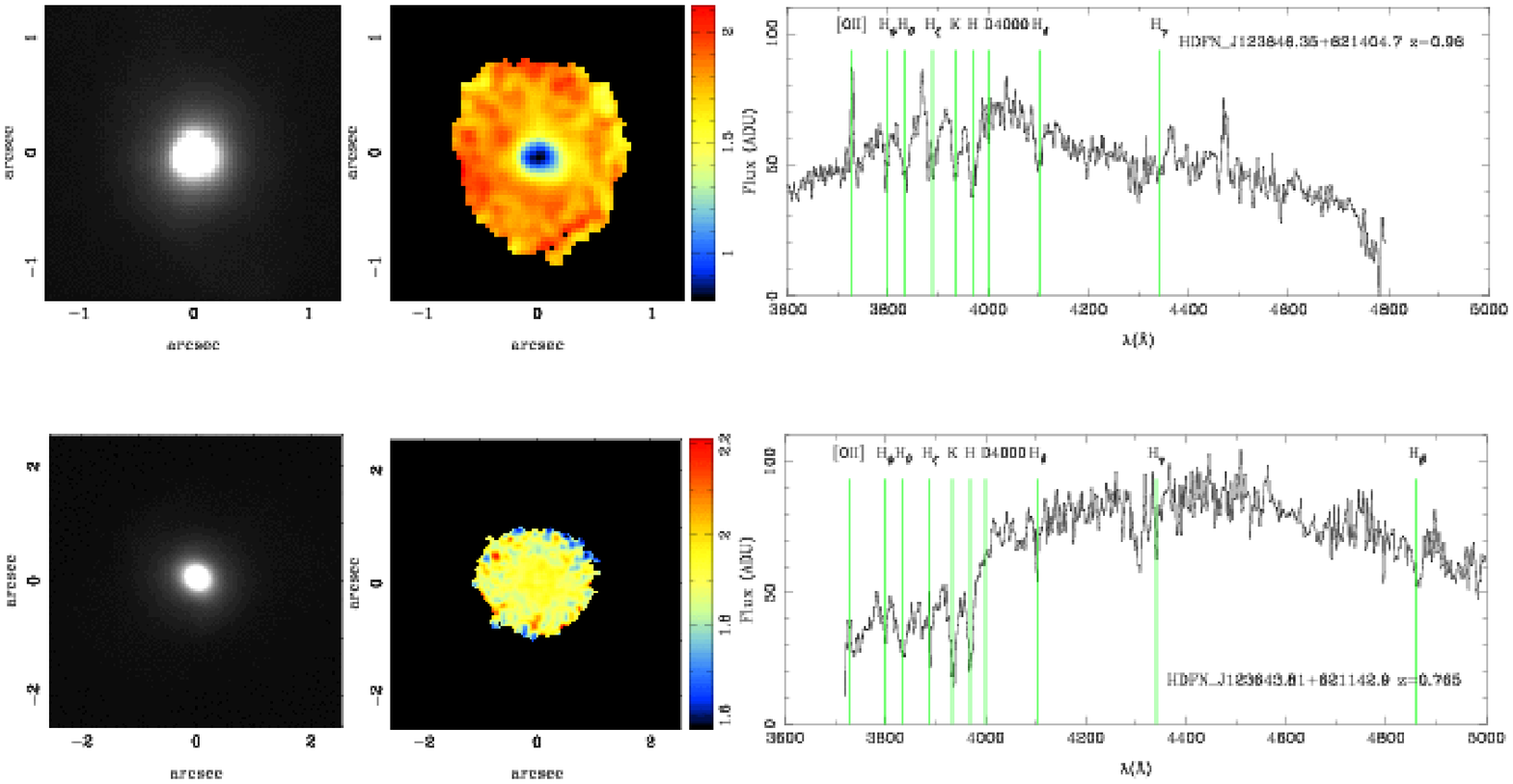,width=14cm}}

\noindent{\bf Figure 11:}\quad {\em Color inhomogeneities in HDF field
ellipticals suggest continued star formation is occurring,
possibly as a result of hierarchical assembly. Each row displays
the $I$ band HST image, a $V-I$ color image and the Keck LRIS spectrum.
The top set refers to a $z$=0.92 elliptical with a blue core; its
spectrum shows features indicative of active star formation ([O
II] emission and deep Balmer absorption lines). The bottom set
refers to a quiescent example at $z$=0.966 whose spectrum is
consistent with an old stellar population. The amount of blue
light can be combined with the depth of the spectral features to
statistically estimate the amount and timescale of recent star
formation.}
\medskip

\end{itemize}

What evolution is found in the properties of other kinds of
galaxy? Brinchmann et al (1998) secured HST images for a sizeable
and statistically-complete subset of CFRS and LDSS redshift survey
galaxies and found the abundance of spirals to $I$=22 - a flux
limit which samples 0.3$<z<$0.8 - is comparable to that expected
on the basis of their local abundance if their disks were somewhat
brighter and bluer in the past as evidenced from surface
photometry (\cite{lilly99}). In practice, however, the
detectability of spiral disks is affected by a number of possible
selection effects (\cite{simard99,bouwens00}) and it may be some
time before a self-consistent picture emerges.

A less controversial result from Brinchmann et al (1998) claimed in
earlier analyses without redshift data
(\cite{glazebrook95,driver95}) is the remarkably high abundance of
morphologically-peculiar galaxies in faint HST data. Brinchmann et
al quantified this in terms of the luminosity density arguing that
a substantial fraction of the claimed decline in the blue
luminosity density since $z\simeq$1 (c.f. Figure 6) arises from
the demise of this population (Figure 12).

\medskip
\centerline{\psfig{file=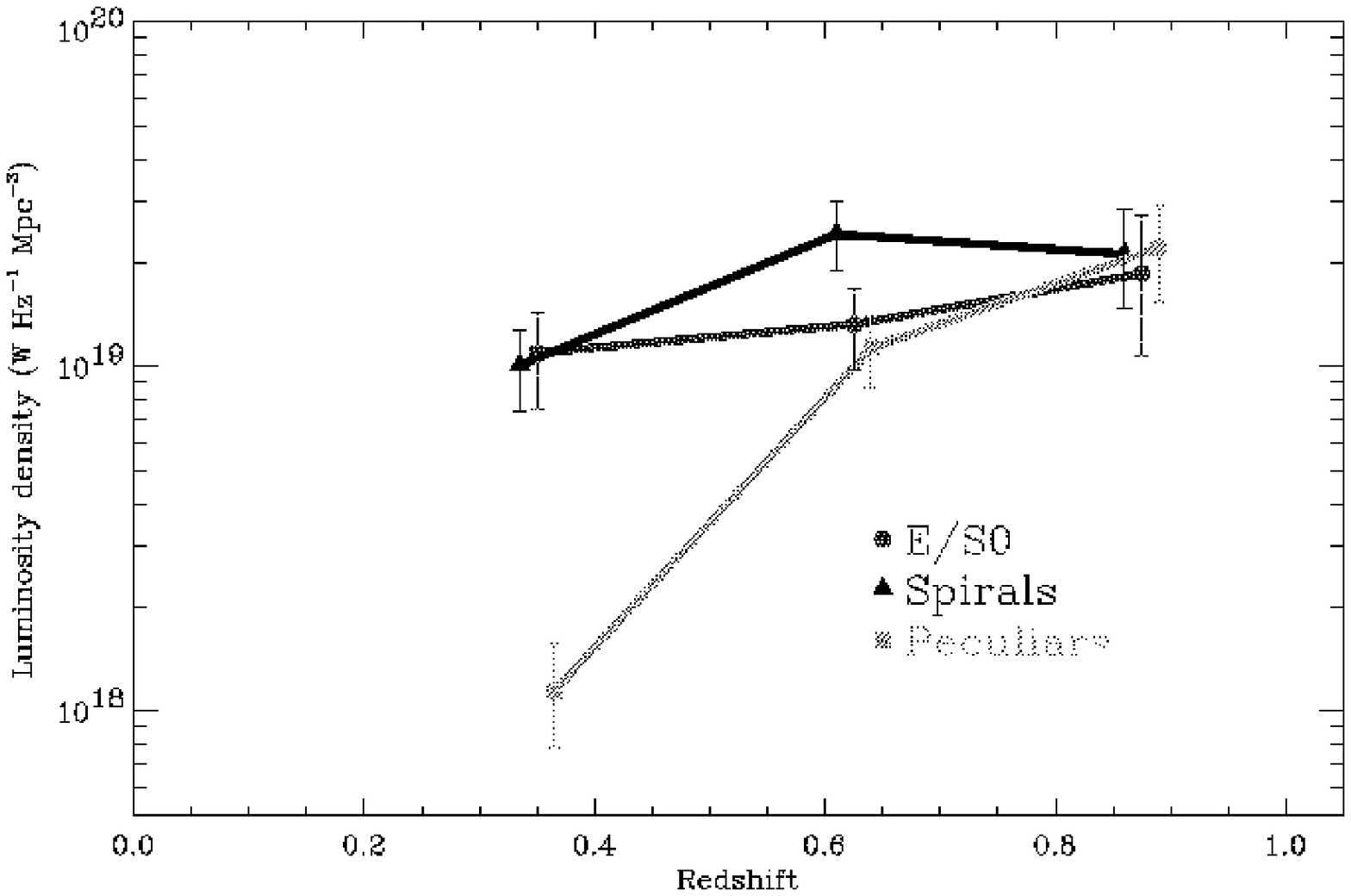,width=11cm}}

\noindent{\bf Figure 12:}\quad {\em The morphological dependence of the
blue luminosity density from that subset of the CFRS/LDSS redshift
survey imaged with HST (Brinchmann et al 1998). The marked decline
in the luminosity density of galaxies with peculiar morphology
over 0$<z<$1 is the primary cause for steep slope in the blue
faint galaxy counts.}
\medskip

Given our earlier concerns with over-interpreting the cosmic star
formation history, should we be cautious in drawing conclusions
from Figure 12? Although Brinchmann et al's redshift sample is
small, the basic result is consistent with the HST morphological
number counts where much larger samples are involved. Whereas
early skeptics argued that morphologically-peculiar galaxies
represent regular systems viewed at unfamiliar ultraviolet
wavelengths, recent NICMOS imaging (see Dickinson's lectures)
suggests such `morphological bandshifting' is only of minor
consequence. In quantitative detail, as before, uncertain
corrections must be made for the effects of the flux limited
sample and of course extinction is a major uncertainty. However,
it seems inescapable that the bulk of the decline in blue light
(the so-called {\em faint blue galaxy problem}, \cite{rse97})
arises from the demise of a population of late-type and
morphologically-peculiar systems. A key question therefore is what
happened to this population? We will address this problem in the
next section.

\section{Constraining the Masses of Distant Galaxies}

A recurring issue arises from the discussions in the earlier
sections. Whilst observers are, with some restrictions, able to
measure distant galaxy properties such as rest-frame colors,
luminosities and star formation rates, these may be poor
indications of the underlying stellar and total masses predicted
most straightforwardly by contemporary models of structure
formation. Either we put our faith in the forward modelling of the
readily-available observables (i.e. we invest a lot of effort in
understanding the complexities of feedback, Figure 7), or we
consider how to measure galactic masses.

Ideally we seek methods for determining the {\it total} mass
(baryonic plus the dark matter halo) but this seems out of reach
for the moment except for local systems with tracers of the larger
halo in which galaxies are thought to reside. Useful tracers here
include the dynamical properties of attendant dwarf galaxies
(\cite{zaritsky98}) and globular clusters (\cite{huchra98}). A
promising route in the future might be galaxy-galaxy gravitational
lensing (\cite{blandford92}). Here a foreground population is
restricted in its selection, perhaps according to morphology or
redshift, and the statistical image distortions in a background
population analysed. Early results were based on HST data, for
cluster spheroidals (\cite{natarajan99}) and various field
populations (\cite{griffiths96}), however with extensive panoramic
data from the Sloan Digital Sky Survey, convincing signals can be
seen with ground-based photometry (\cite{fischer00}). Again,
photometric redshifts will be helpful in refining the sample
selection and in determining the precise redshift distribution
essential for accurate measures on an absolute scale.

Unfortunately, promising though the technique appears, the
restrictions of galaxy-galaxy lensing are numerous. It only gives
mass estimates for statistical samples: the signal is too weak to
be detected in individual cases, unless a strong lensing feature
is seen (\cite{hogg96}). Moreover, the redshift range and physical
scale on which the mass is determined is defined entirely by
geometrical factors and, ultimately, one may never be able to
apply the method to galaxies beyond $z\simeq$1.

Extensive dynamical data is becoming available for restricted
classes of high redshift galaxy, via linewidth measures
(\cite{koo95}), resolved rotation curves (\cite{vogt97}) for
sources with detectable [O II] 3727 \AA\ emission, and via
internal stellar velocity dispersions for absorption line galaxies
such as spheroidals. Under certain assumptions, these give mass
estimates and have enabled the construction of the fundamental
plane for distant spheroidals (\cite{treu00}) and the Tully-Fisher
relation for high redshift disk galaxies (\cite{vogt97}). The
greatest progress in the former has been in constructing the
fundamental plane in rich clusters (\cite{vandokkum98}) where slow
evolution in the inferred mass/light ratio for cluster ellipticals
is consistent with a high redshift of formation (see $\S$5). For
the emission line studies it is not straightforward to convert
data obtained over a limited spatial extent into reliable masses
even for regular well-ordered systems. For compact and irregular
sources, the required emission lines may come from
unrepresentative components yielding poor mass estimates
(\cite{lehnert96}).

The prospects improve significantly if we drop the requirement to
measure the {\em total} mass and are willing to consider only the
{\em stellar mass}. In this case, the near-infrared luminosity is
of particular importance. Broadhurst et al (1992) and Kauffmann \&
Charlot (1998) have demonstrated that the $K$ (2$\mu$m) luminosity
is a good measure of its underlying stellar mass {\em regardless
of how that mass assembled itself} (Figure 13). This remarkable
fact arises because $K$-band light in all stellar populations
(whether induced in bursts or continuous periods of activity)
arises from long-lived giants whose collective output mirrors the
{\em amount} of past activity, smoothing over its production
timetable.

\medskip
\centerline{\psfig{file=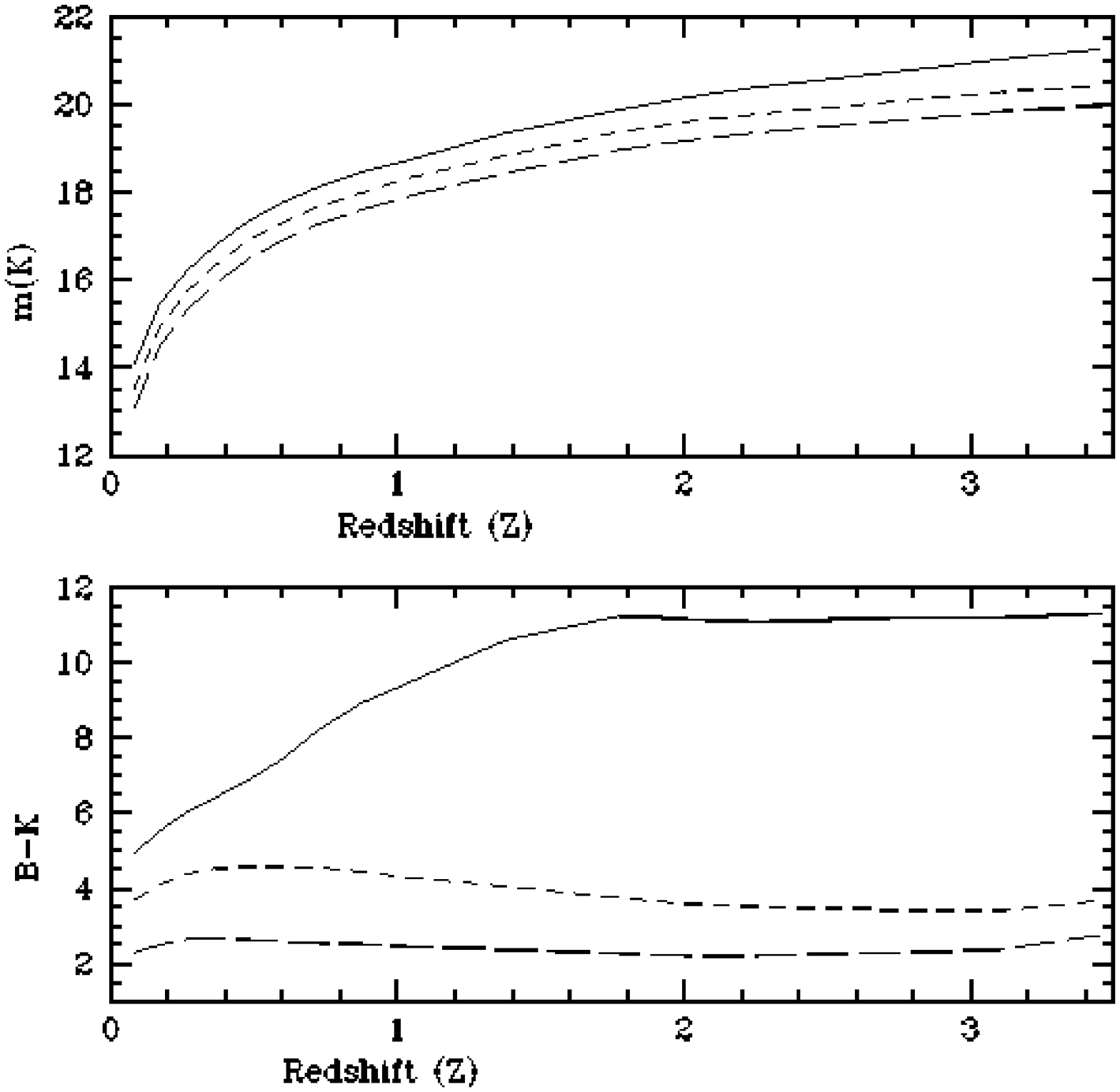,width=11cm}}

\noindent{\bf Figure 13:}\quad {\em The $K$-band luminosity is a good
measure of the underlying stellar mass irrespective the past star
formation history (Kauffmann \& Charlot 1998). The curves show the
observed $K$ magnitude as a function of the redshift at which such
an object is selected for a system containing 10$^{11}M_{\odot}$
produced according to a variety of star formation histories. Even
across extreme cases (single burst at $z=\infty$, solid line, to a
constant star formation rate to the epoch of observation,
short-dash), the $K$-band output remains the same to within a
factor of $\simeq$2. The lower panel shows, how different the
observed optical-infrared color would be in these cases.}
\medskip

A deep K-band redshift survey thus probes the very existence of
massive systems at early times. A slightly incomplete survey to
$K$=20 (\cite{cowie96}) and a complete photometric survey to
$K$=21 (\cite{fontana99}) indicates an apparently shortfall of
luminous $K$ objects beyond $z\simeq$1-1.5 c.f. pure luminosity
evolution models. Unfortunately, small sample sizes,
field-to-field clustering, spectroscopic incompleteness and
untested photometric redshift techniques beyond z$\simeq$1 each
weaken this potentially important conclusion. An important goal in
the immediate future must be to reconcile these claims with the
apparently abundant (and hence conflicting) population of
optical-infrared red objects to $K\simeq$19-20
(\cite{daddi00,mccarthy00}).

\medskip
\begin{figure}
  \begin{tabular}{c}
\centerline{\psfig{file=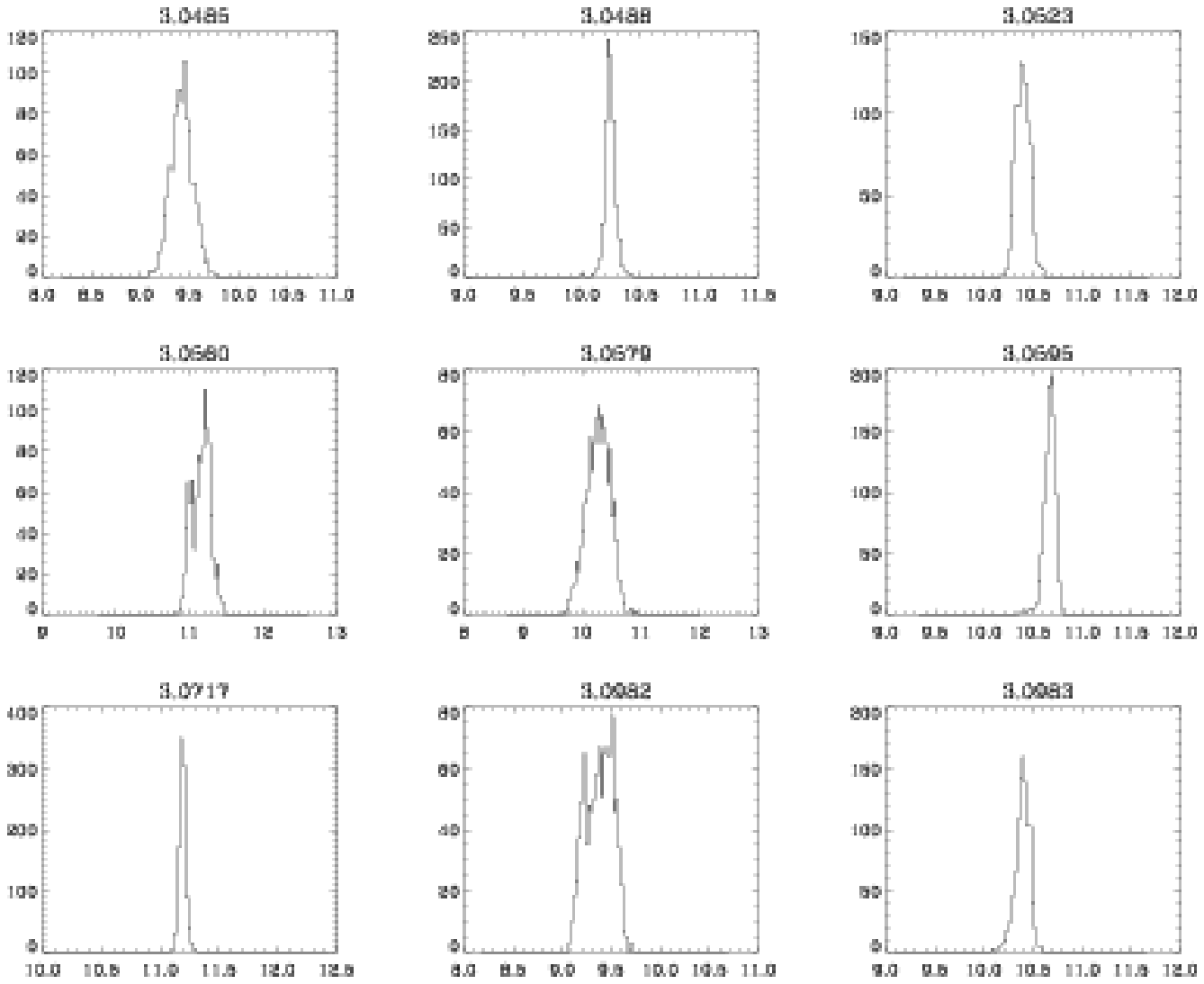,width=8cm}} \\   
\centerline{\psfig{file=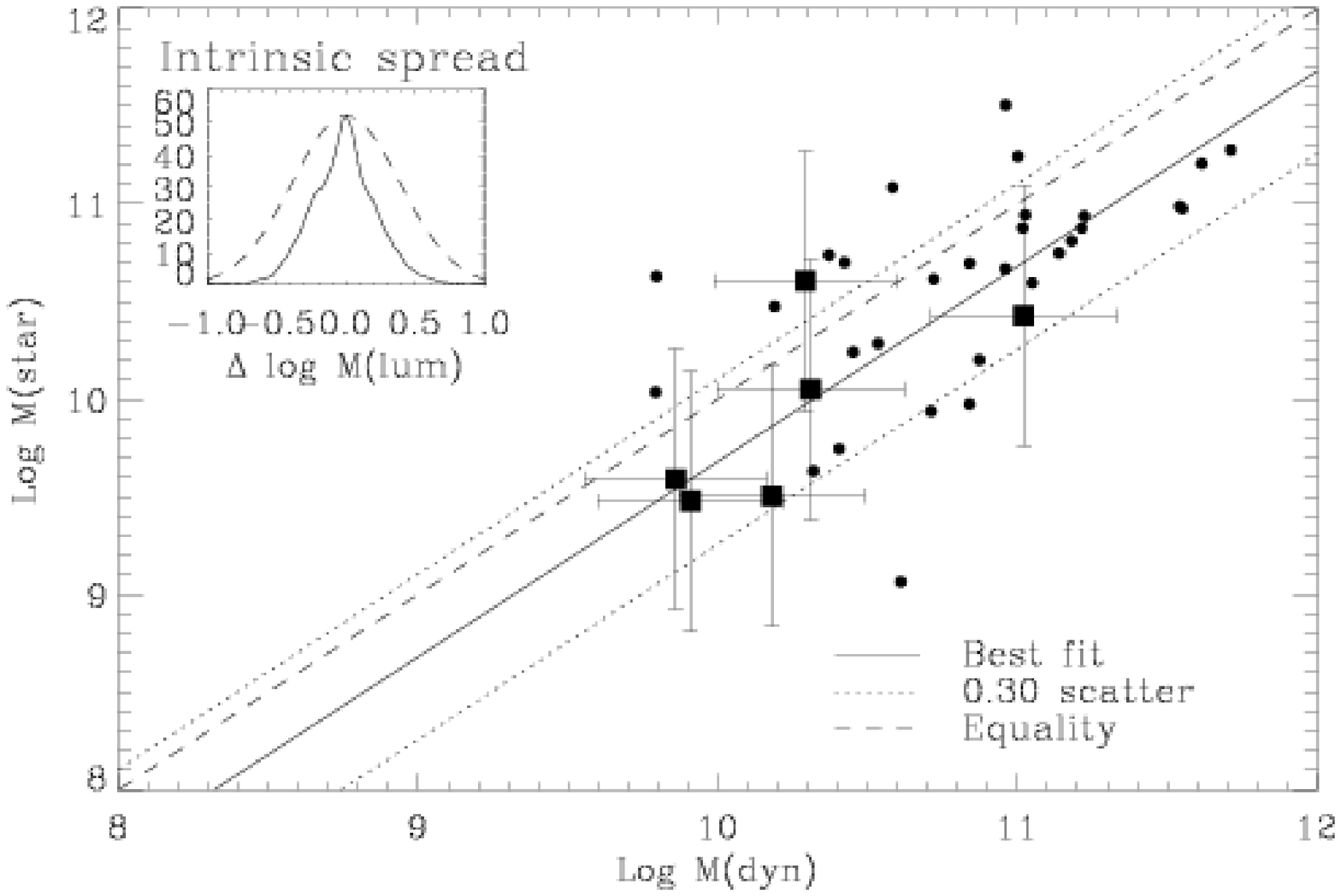,width=11cm}} \\
\end{tabular}
\noindent{\bf Figure 14:}\quad {\em The infrared method for determining the
stellar mass of a distant galaxy (\cite{rse00}). The technique
fits the observed SED for a galaxy of known redshift in the
context of evolutionary synthesis models where the stellar mass is
the fitted variable. (Top) Likelihood functions for the derived
logarithmic stellar mass for sample galaxies in the CFHT/LDSS
redshift survey (Brinchmann, Ph.D. thesis 1998); a typical
uncertainty of 30-50\% is secured at I$\simeq$22. (Bottom)
Correlations of stellar and dynamical mass for both low z (circles) and 
high z (squares with error bars) galaxies from the analysis of 
Brinchmann \& Ellis (2000).}
\end{figure}
\medskip

The precision of the technique introduced by Kauffmann \& Charlot
(1998) can be improved if the optical-infrared color is available
as an extra parameter (\cite{rse00}). In this way a first-order
correction can be made for the past star formation history and
hence the effect of the spread in the lower panel of Figure 13 can
be used to improve the mass estimate. Importantly, such a
technique for determine accurate stellar masses can then be
applied to {\em all} galaxies, regular or peculiar, irrespective
of their dynamical state and over a range in redshift (providing
the data is sufficiently precise). The technique can be considered
as a modification of that frequently utilised in estimating
photometric redshifts. The observed optical-infrared SED for an
object of known redshift is used to optimally fit the {\em stellar
mass} rather than the redshift in the framework of an evolutionary
synthesis code. Stellar masses can be derived to within a random
uncertainty of 30-50\% by this technique although at present there
is no reliable way to verify the results except by comparison with
independent dynamical measures (Figure 14).

\bigskip
\centerline{\psfig{file=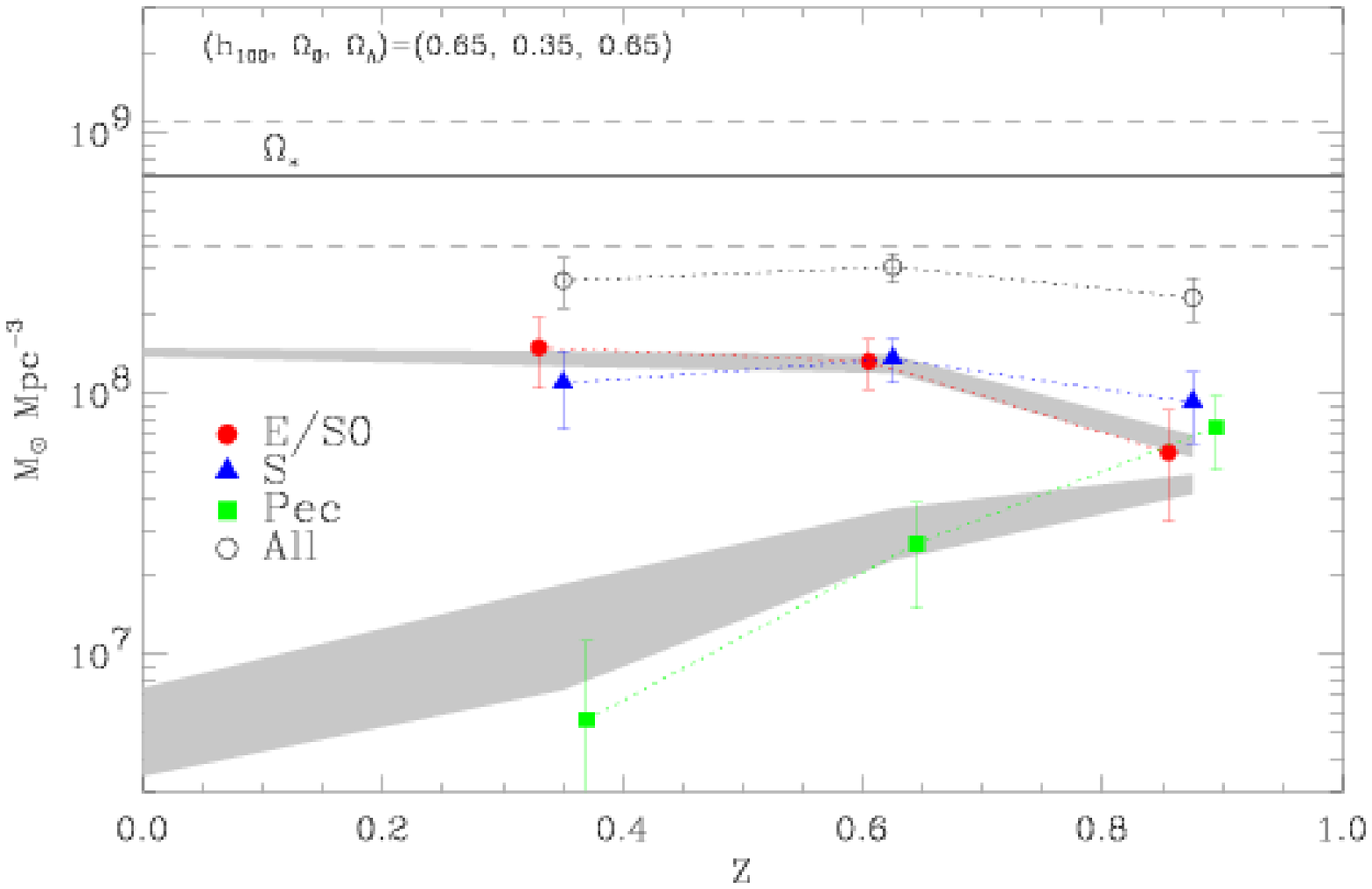,width=11cm}}

\noindent{\bf Figure 15:}\quad {\em Evolution of the stellar mass density
$\rho_{stars}(z,T)$ from the analysis of Brinchmann \& Ellis
(2000). A remarkable decline with time in stellar mass density is
seen for the morphologically-peculiar class which argues against a
truncation of their star formation activity as the primary cause
for their demise. Brinchmann \& Ellis argue that this population
must be transforming, possibly via mergers, into the regular
classes. A simple model which implements a likely
redshift-dependent merger rate (\cite{lefevre00}) with elliptical
products can broadly reproduce the trends observed (shaded area of
the plot).}

\bigskip
\section{Origin of the Hubble Sequence}
\medskip

The availability of stellar masses for {\em all} types enables the
construction of a powerful evolutionary plot, analogous to Figure
6, involving the {\em stellar mass density}, {$\rho_{stars}(z,T)$,
as a function of morphology $T$. Whilst the stellar mass density
can {\em grow} by continued star formation, unlike the {\em UV
luminosity density}, $\rho_{UV}$, it is difficult to imagine how
it can {\em decline}. As we saw earlier $\rho_{UV}$ can decline
significantly in only 1-2 Gyr because of an abrupt truncation of
activity. However, such a change would have very little effect on
the infrared output as illustrated in Figure 13.

Brinchmann and Ellis (2000) secured $K$ luminosities and
optical-IR SEDS for over 300 galaxies in the CFRS/LDSS and Hawaii
survey fields and derive $\rho_{stars}(z,T)$ (Figure 15).
Estimating the integrated stellar mass density is prone to all of
the difficulties reviewed earlier for the luminosity density and
there is the added complication that the redshift surveys in
question are {\em optically-selected} and thus must miss some
(red) fraction of a true $K$-limited sample. Accordingly, the mass
densities derived are lower limits to the true values.

Remarkably, $\rho_{stars}(z,T)$ is a declining function for the
intriguing population of morphologically-peculiar galaxies.
Whereas the declining UV luminosity density could imply a fading
population, such an explanation cannot be consistent with Figure
15 which argues, instead, that the objects are genuinely
disappearing into other systems. The most logical explanation for
their declining contribution to the stellar mass density is that
morphologically-peculiar objects are being transformed, e.g. by
mergers, into regular objects.

\medskip
\centerline{\psfig{file=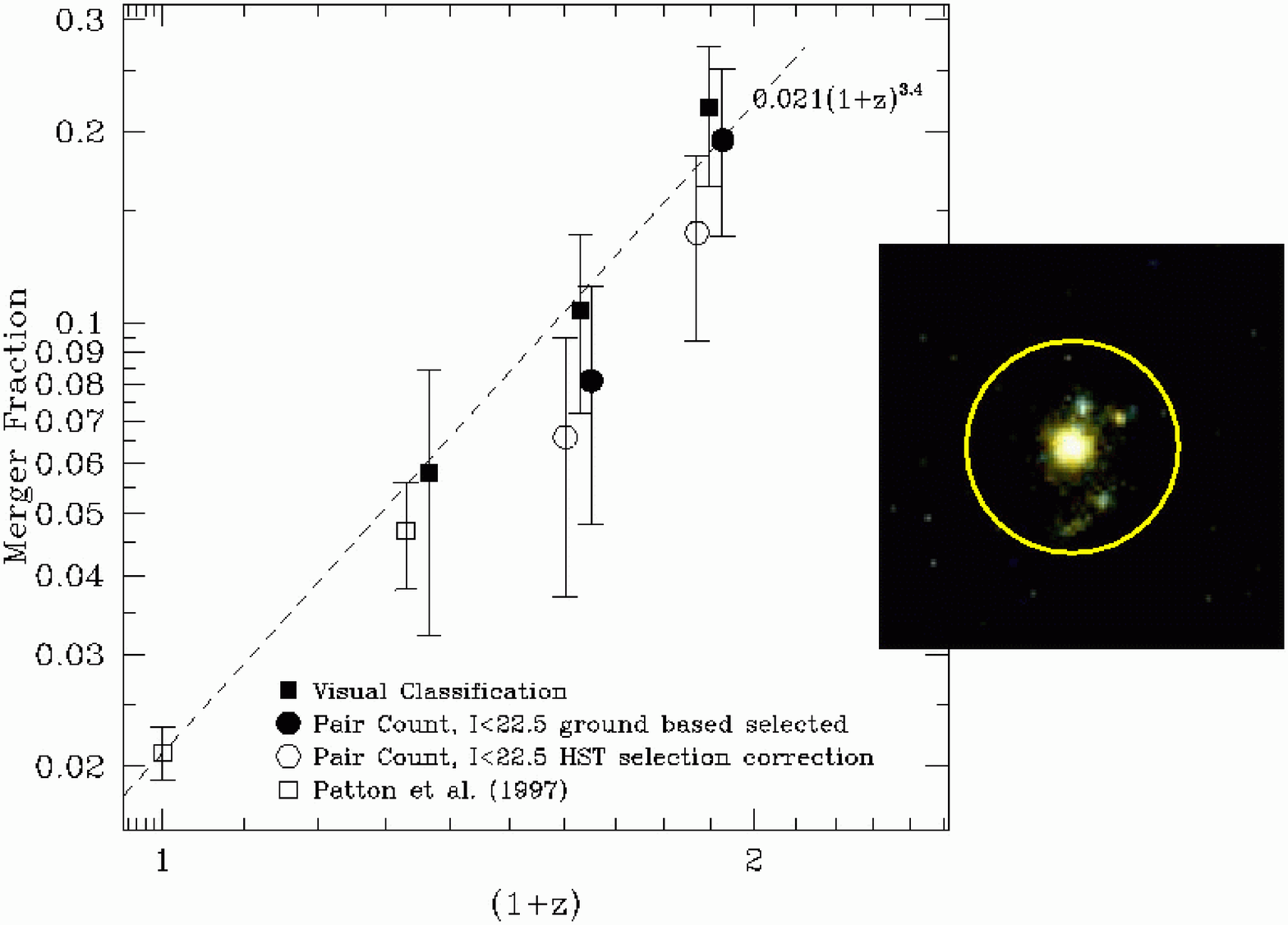,width=11cm}}

\noindent{\bf Figure 16:}\quad {\em An increase in the merger {\em
fraction} as a function of redshift from the HST analysis of
LeFevre et al (2000). Galaxies of known redshift were examined for
satellites brighter than a fixed rest-frame luminosity within a
projected radius of 20$h^{-1}$ kpc and corrections made for
unrelated line-of-sight contamination. This redshift-dependent
merger {\em rate} was adopted by Brinchmann \& Ellis (2000) in
Figure 15.}
\medskip

Merging has been an attractive means for governing the evolution of
galaxies for many years (\cite{toomre72,rocca89,beg92}) and of
course is fundamental to the hierarchical formation picture.
However it has been extremely difficult to determine the observed
rate at intermediate redshift. The fundamental problem is that we
observe galaxies at various look-back times via discrete
`snapshots' without ever being able to {\em prove} two associated
systems are destined to merge on a particular timescale. Using the
CFRS/LDSS HST dataset referred to earlier, LeFevre et al (2000)
undertook a quantitative survey of the {\em fraction} of luminous
galaxies with satellites brighter than a fixed absolute magnitude
within a 20$h^{-1}$ kpc metric radius and, after allowance for
projection effects, determined the merger {\em fraction} increases
with redshift as $\propto \;(1\,+\,z)^{3.4 \pm 0.5}$ - a result
consistent with earlier ground-based efforts. Sadly, it is not
straightforward to convert the proportion of galaxies with
associated sources into a physical merger rate or, as ideally
required, a mass assembly rate without some indication of the
dynamical timescale for each merger and the mass of each
satellite. Moreover, there are several annoying biases that affect
even the derived merger fraction.

Brinchmann \& Ellis (2000) attempted to reconcile the decline of
the morphologically-peculiar population, the redshift dependence
of the LeFevre et al merger fraction and associated evidence for
continued formation of ellipticals (\cite{menanteau00}) into a
simple self-consistent picture. They transferred the dominant
population of morphologically-irregular galaxies, via the
$z$-dependent merger rate, into a growth in the regular galaxies
(shaded area of Figure 15). This is clearly a simplistic view, but
nonetheless, gives a crude empirical rate at which regular galaxies
are assembling. If correct, how does this agree with mass assembly
histories predicted, say in $\Lambda$CDM?

\medskip
\centerline{\psfig{file=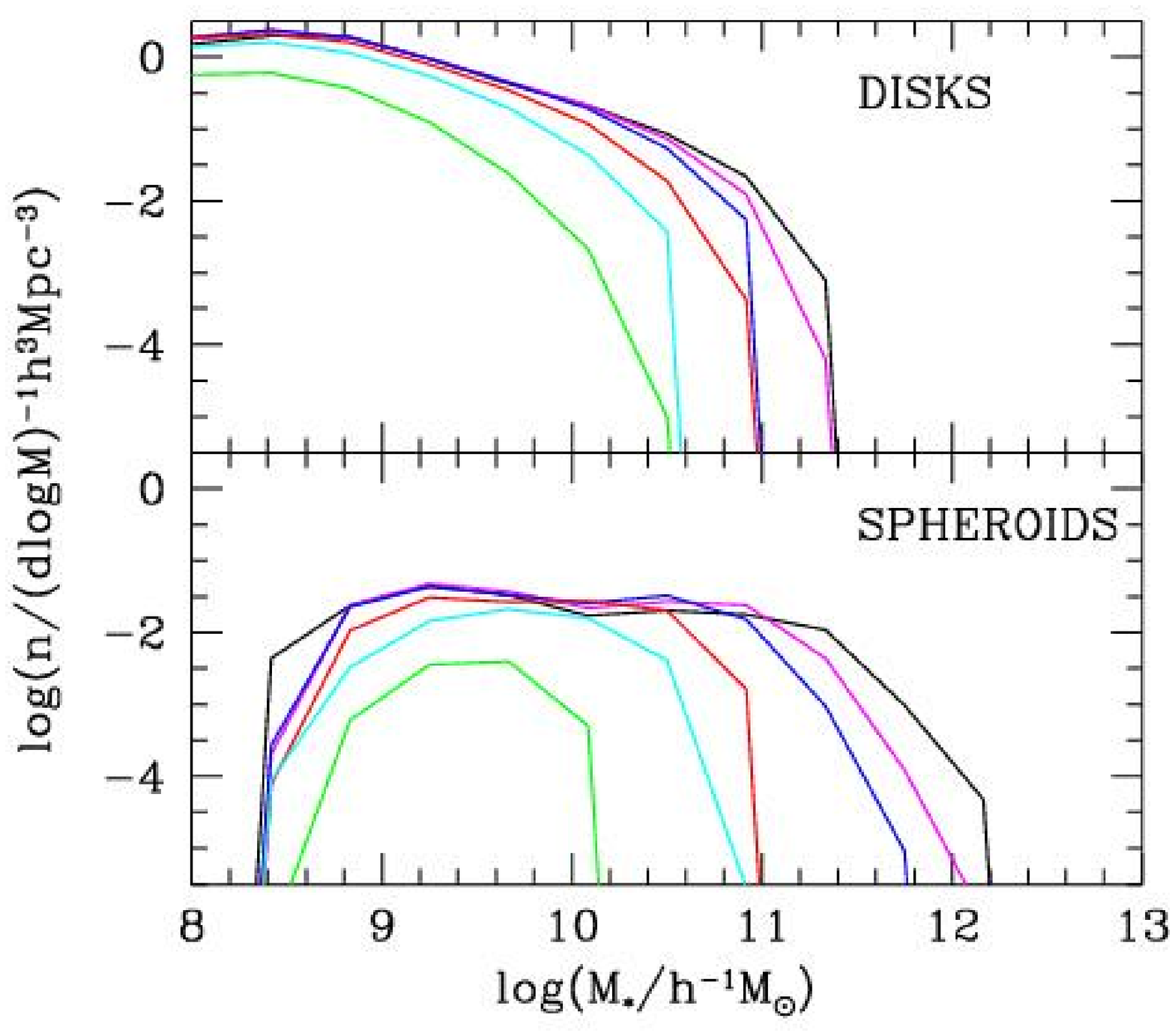,width=10cm}}
\noindent{\bf Figure 17:}\quad {\em Predicted evolution in stellar mass
functions for disk and spheroidal populations in a $\Lambda$CDM
hierarchical model (Frenk, priv. comm). The curves define mass
functions as a function of redshift ($z$=0,0.5,1,2, from right to
left). Modest growth over 0$<z<$2 is expected for disk galaxies
but significant growth is predicted for massive spheroidals.}
\medskip

Figure 17 shows a recent prediction of the assembly history of
spheroids and disks (Frenk, private communication). Although there
are some discrepancies between this and its equivalent prediction
from Kauffmann \& Charlot (1998, Figure 3), the trends are clear.
The strongest evolutionary signal is expected in terms of a recent
assembly of massive spheroids; the equivalent growth rate in
stellar disks is more modest. To the extent it is currently
possible to test this picture, the qualitative trend is supported
by the data. Field ellipticals are certainly still assembling
(\cite{menanteau00}) but perhaps more slowly than expected
according to Figure 17; unfortunately deeper samples with
redshifts are needed for a precise statement. Brinchmann (in
prep.) has examined the stellar mass growth rate in disks using
the infrared-based method over 0$<z<$1 and finds only modest
changes. This is very much a developing area and one that would
benefit from significantly enlarged HST datasets chosen to overlap
the growing faint redshift survey databases.

\medskip
\centerline{\psfig{file=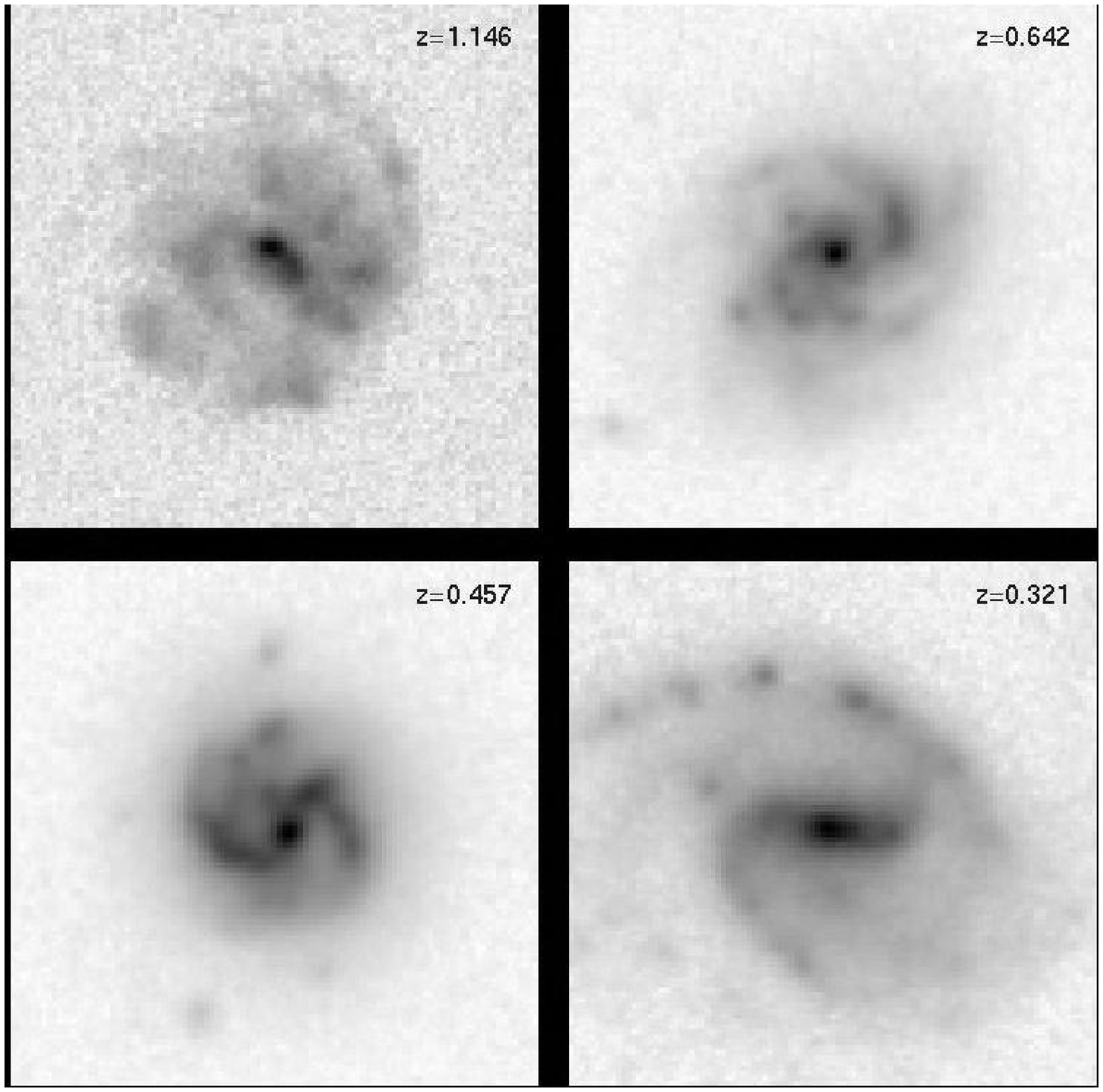,width=10cm}}

\noindent{\bf Figure 18:}\quad {\em Face-on barred spirals of known
redshift in the Hubble Deep Field. Abraham et al (1999) claim that
all such systems should easily be recognised to z$\simeq$1 in
HDF-quality data whereas, beyond z$\simeq$0.6, there appears to be
a marginal paucity of such systems compared to their non-barred
counterparts. If supported by further data, this could indicate an
epoch corresponding to the `dynamical maturing' of stellar disks.}
\medskip

The HST data, particularly that in the Hubble Deep Fields (HDF),
is an astonishingly rich resource which is still not completely
exploited. As an indication of what might be possible with future
instrumentation, I will close with some remarks on the important
role that bulges and bars may play in the history of the Hubble
sequence.

50\% of local spirals have bars which are thought to originate
through dynamical instabilities in well-established
differentially-rotating stellar disks. If we could determine the
epoch at which bars begin appearing, conceivably this would shed
some light on how recently mature spirals came to be. Via careful
simulations based on local examples, Abraham et al (1999) showed
that face-on barred galaxies should be recognisable to $z\simeq$1
in the HDF exposures. In fact, many are seen (Figure 18) but
tantalisingly the barred fraction of face-on spirals appears to
drop beyond a redshift z$\simeq$0.6. The effect is marginal but
illustrative of a powerful future use of morphological imaging
with the Advanced Camera for Surveys.

\medskip
\begin{figure}
\centerline{\hbox{
\psfig{file=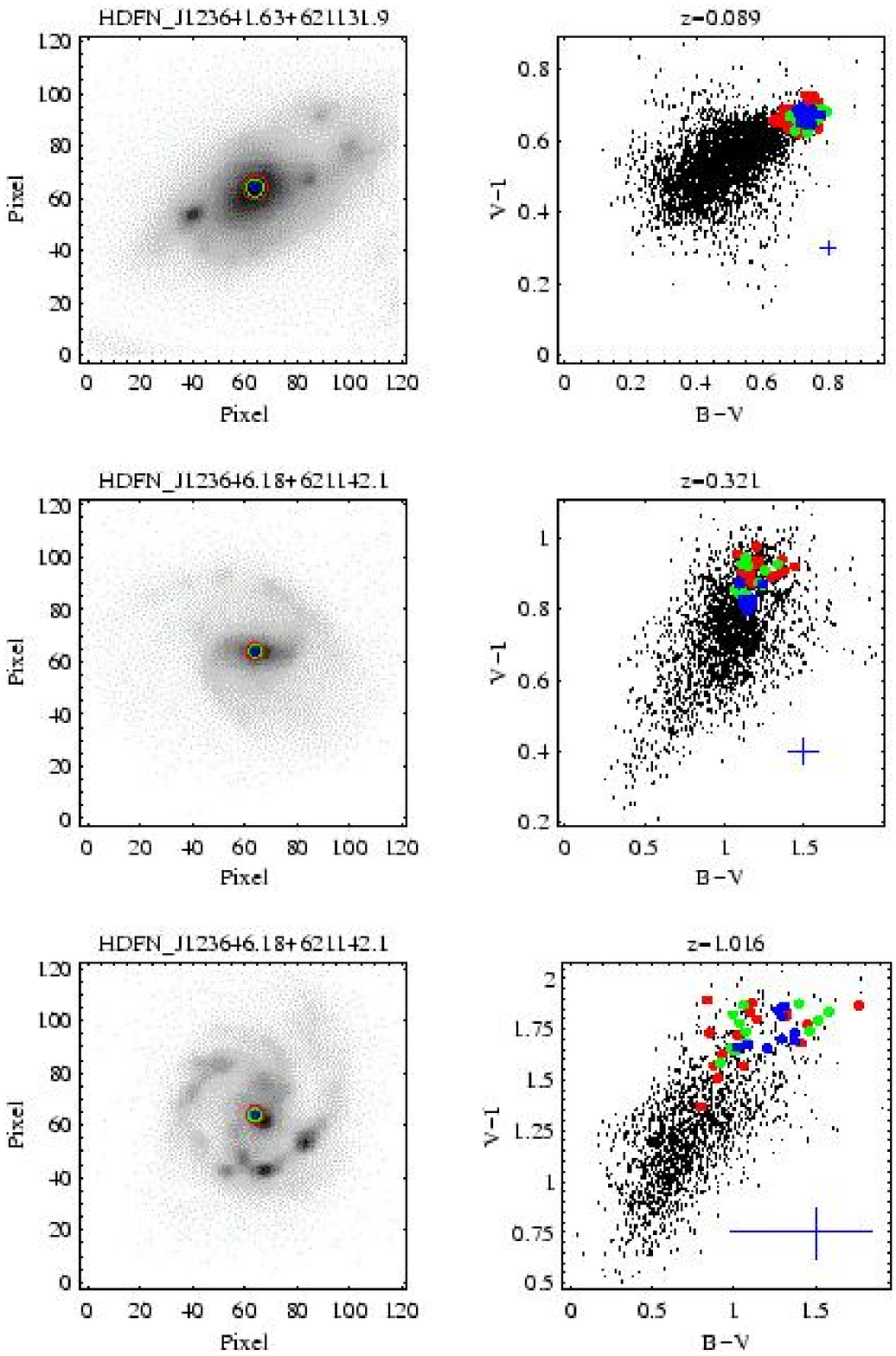,width=7cm}    
\psfig{file=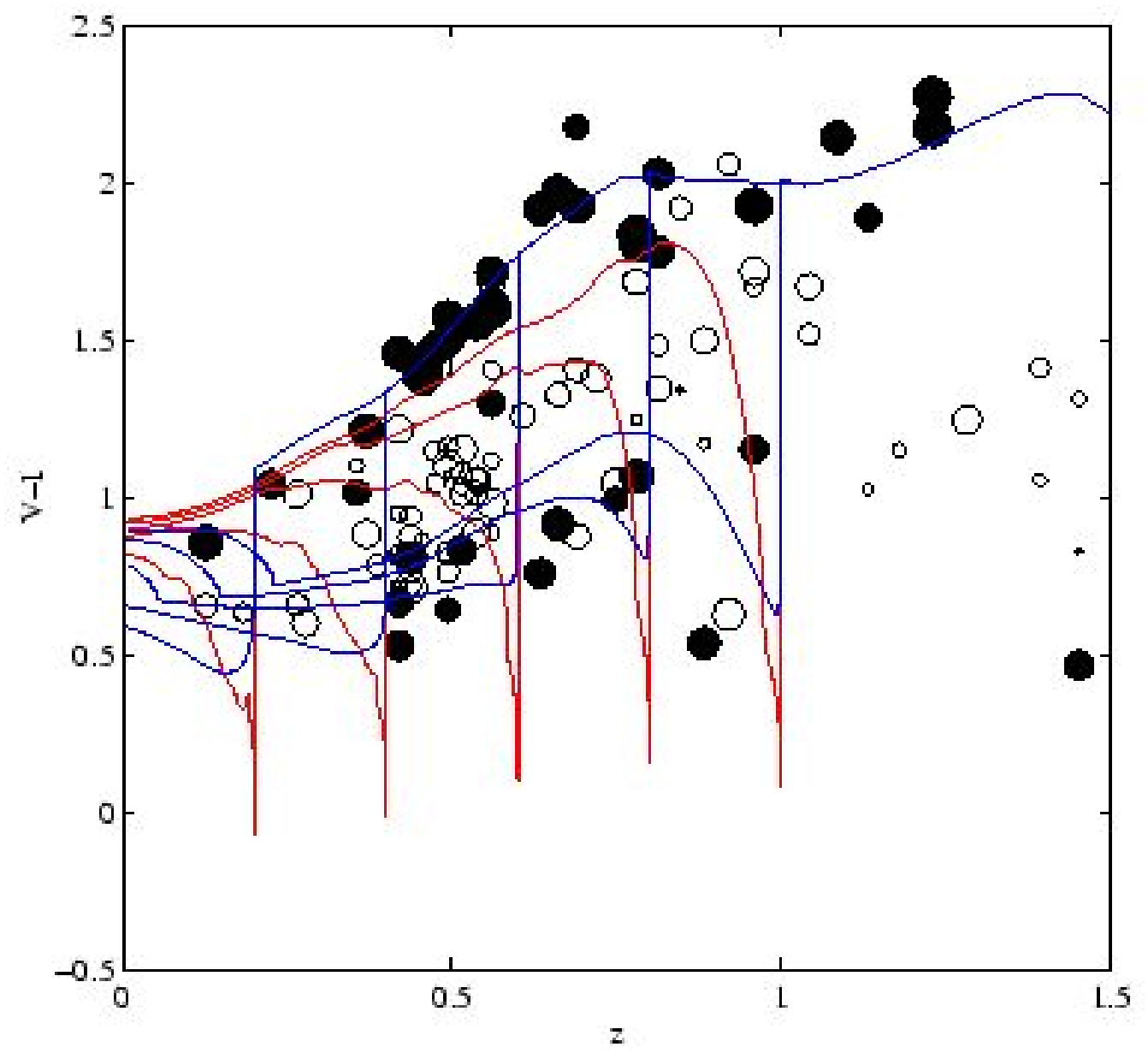,width=6.5cm}
}}
\noindent{\bf Figure 19:}{\em The remarkable diversity of
intermediate redshift spiral bulges in the Hubble Deep Fields as
revealed in the analysis of Ellis et al (2000). (Top) Selected
face-on spirals in the HDF with pixel-by-pixel $BVI$ color
distributions. The marked points represent various aperture
selections which serve to define the mean bulge color; in each
case the bulge remains the reddest part of the spiral galaxies.
(bottom) $V-I$ aperture color for bulges (open circles) and
integrated color for ellipticals (filled circles) versus redshift.
Bulges are generally more diverse with a mean color bluer than
their elliptical counterparts. Curves illustrate that a continued
infall of 5\% by mass over 1-2 Gyr timescales could explain the
observed trends.}
\label{}
\end{figure}
\medskip

The story with bulges is also unclear, although potentially
equally exciting. Traditionally, bulges were thought to represent
miniature ellipticals which formed monolithically at high redshift
(\cite{els62}). Detailed studies of local examples, including the
Galactic bulge, have shown a considerable diversity in properties,
both in integrated color and even in their photometric structure
(\cite{wyse99}). There is some evidence of a bimodality in the
population; prominent bulges in early type spirals share surface
brightness characteristics of ellipticals, whereas those in
late-type spirals are closer to exponential disks. This might
indicate two formation mechanisms, one primordial (as in the
traditional picture), the other related perhaps to the merging
assembly history or via disk instabilities through what is termed
`secular' evolution.

Taking advantage of the HDF images, including those from NICMOS,
Ellis et al (2000) have examined the color distribution for a
large sample of spirals bulges of known redshift and compared
these colors with their integrated equivalent for the HDF
ellipticals. If bulges are miniature ellipticals formed at high
redshift, one would expect similar trends. Interestingly, in the
hierarchical picture, one expects bulges to be {\em older} and
presumably redder than ellipticals (since the latter predominantly
form from merged disk systems which most likely contain bulges as
early merger remnants). Ellis et al (2000) find intermediate
redshift bulges are the reddest part of a typical spiral but,
surprisingly, they are often bluer than their elliptical
counterparts and far less homogeneous as a population.
Contamination from disk light is an obvious concern though
simulations suggest only modest bias arises to redshifts where
these trends become prominent. What could be responsible for this
puzzling behavior? Evolutionary synthesis modelling suggest only a
modest amount of star formation corresponding to continued infall
of $\simeq$5\% by mass would be needed to explain the bluing.

\section{Conclusions}
\medskip

In summary, despite the frantic increase in publication rate in
this field, there is an enormous amount of work still to be done,
both observationally, in exploiting the connection between
resolved images from HST and ground-based spectroscopy, and
theoretically, in predicting more accurately the expected
evolutionary histories of resolved components. In my opinion the
subject suffers too much from a satisfaction with simply
replicating, according to a particular theory, a range of
observations. This is particularly dangerous when the observables
are luminosities, colors and star formation rates since the
theoretical parameters involved are numerous. The challenge will be to overcome
the obvious limitations we presently face in determining galactic
masses for complete samples of galaxies viewed at various
look-back times, as well as integrating the growing body of data
being obtained in the far infrared and sub-mm spectral regions.

\begin{acknowledgments}
I thank my students, past and present, and collaborators at
Cambridge, Caltech and elsewhere for allowing me to present the
results of unpublished work undertaken with them. I also thank
Marc Balcells, Ismael Perez-Fournon and Francisco Sanchez  for
inviting me to Tenerife to give these lectures and for their
remarkable patience in waiting for this written version.
\end{acknowledgments}

\end{document}